\documentclass[]{JHEP3} 

\usepackage{epsfig,multicol}

\usepackage{graphicx}
\usepackage{slashbox}

\newcommand\fverb{\setbox\pippobox=\hbox\bgroup\verb}
\newcommand\fverbdo{\egroup\medskip\noindent%
            \fbox{\unhbox\pippobox}\ }
\newcommand\fverbit{\egroup\item[\fbox{\unhbox\pippobox}]}
\newbox\pippobox

\newcommand{\bastar}{\begin{eqnarray*}}
\newcommand{\eastar}{\end{eqnarray*}}
\newskip\humongous \humongous=0pt plus 1000pt minus 1000pt

\relax
\newcommand{\be}{\begin{equation}}
\newcommand{\ee}{\end{equation}}
\newcommand{\bea}{\begin{eqnarray}}
\newcommand{\eea}{\end{eqnarray}}
\newcommand{\X}{{\vec X}}
\newcommand{\pro}{\partial}
\newcommand{\n}{\hat n}
\newcommand{\oneg}{\displaystyle\frac{1}{g}}

\newcommand{\D}{{\hat D}}

\newcommand{\A}{{\vec A}}
\newcommand{\valpha}{{\vec \alpha}}

\newcommand{\dfrac}{\displaystyle\frac}
\newcommand{\ba}{\begin{array}}
\newcommand{\ea}{\end{array}}

\newcommand{\nn}{\nonumber}
\newcommand{\hn}{\hat n}

\title{New Interpretation of Skyrme Theory}

\author{Y. M. Cho \\
C. N. Yang Institute
for Theoretical Physics, \\
State University of New York, Stony Brook, New York 11794, USA \\
and \\
School of Physics, College of Natural Sciences, \\
Seoul National University, Seoul 151-742, Korea \\
E-mail: \email{ymcho@yongmin.snu.ac.kr}}

\author{B. S. Park \\ School of Physics,
College of Natural Sciences, \\
Seoul National University, Seoul 151-742, Korea \\
E-mail: \email{psychist@phya.snu.ac.kr}}

\author{P. M. Zhang \\ School of Physics,
College of Natural Sciences, \\
Seoul National University, Seoul 151-742, Korea \\
E-mail: \email{zhpm@phya.snu.ac.kr}}

\abstract{
Based on the proposal that the Skyrme theory is a theory of
monopole we provide a new interpretation of Skyrme theory, that
the theory can also be viewed as an effective theory of strong
interaction which is dual to QCD, where the monopoles (not the
quarks) are confined through the Meissner effect. This dual
picture leads us to predict the existence of a topological
glueball in QCD, a chromoelectric knot which is dual to the
chromomagnetic Faddeev-Niemi knot in Skyrme theory, whose mass and
decay width are estimated to be around $60~GeV$ and $8~GeV$. As
importantly, the existence of the magnetic vortex and the magnetic
vortex ring in Skyrme theory strongly indicates that the theory
could also be interpreted to describe a very interesting low
energy condensed matter physics in a completely different
environment. These new interpretations of Skyrme theory puts the
theory in a totally new perspective. }


\begin{document}

\section{Introduction}

The Skyrme theory has played an important role in physics, in
particular in nuclear physics as a successful effective field
theory of strong interaction \cite{skyr,prep,rho,witt}. This is
based on the fact that the theory can be viewed as a non-linear
sigma model which describes the pion physics. I this view
the baryon is identified as a topological soliton made of pions.
The purpose of this
paper is to argue that the theory allows totally different
interpretations which put it in a new perspective. Based on the
proposal that the Skyrme theory is a theory of monopole
\cite{cho01,cho3}, we argue that the theory can also be viewed as
an effective theory of strong interaction which is dual to Quantum
Chromodynamics (QCD). More precisely we argue that the theory is
an effective theory of confinement in which the monopoles, not the
quarks, in QCD are confined. This provides a new interpretation of
Skyrme theory which is orthogonal but complementary to the popular
view. Furthermore, based on the fact that the Skyrme theory has a
built-in Meissner effect which confines the magnetic flux of
monopoles, we also argue that the theory can describe a very
interesting low energy condensed matter physics in a completely
different environment.

A remarkable feature of Skyrme theory is its rich topological
structure \cite{cho01,cho3}. It has been well-known that the
theory allows the skyrmion and the baby skyrmion \cite{skyr,piet}.
But recently it has been shown that it also allows the helical
baby skyrmion and the Faddeev-Niemi knot \cite{cho01,cho3,fadd1}.
More importantly, it contains (singular) monopoles which play a
fundamental role. In fact all the finite energy topological
objects in the theory could be viewed either as dressed monopoles
or as confined magnetic flux of the monopole-antimonopole pair.
For example, the skyrmion can be viewed as a dressed monopole, and
the baby skyrmion can be viewed as a magnetic vortex created by a
monopole-antimonopole pair infinitely separated apart. These
observations have led us to propose that the theory can be
interpreted as a theory of monopole, in which the
monopole-antimonopole pairs are confined to make finite energy
bound states \cite{cho01,cho3}.

In fact the Skyrme theory is a theory of confinement with a
built-in Meissner effect, where the confinement is manifest
already at the classical level. It is this Meissner effect which
allows us to have the baby skyrmion, which is nothing but the
confined magnetic flux of a monopole-antimonopole pair infinitely
separated apart. This confinement mechanism allows us to construct
more interesting topological objects, the helical baby skyrmion
and the Faddeev-Niemi knot. The helical baby skyrmion is a twisted
chromomagnetic vortex which is periodic in $z$-coordinate. The
importance of the helical baby skyrmion is that it assures the
existence of the Faddeev-Niemi knot. This is because the
Faddeev-Niemi knot can be viewed as a vortex ring made of the
helical baby skyrmion, with periodic ends connected together
\cite{cho3,fadd1}. This allows us to interpret the knot as two
magnetic flux rings linked together, the first one winding the
second $m$ times and the second one winding the first $n$ times.

In this paper we present a consistent knot ansatz and construct a
knot solution numerically. Our solution confirms that the knot can
indeed be interpreted as two magnetic flux rings linked together,
whose quantum number is given by the product of two magnetic flux
quanta $mn$, the linking number of two flux rings. This tells that
the knot manifests its topology even at the dynamical level, as
the linking of two magnetic flux rings. This dynamical
manifestation of knot topology assures a supercurrent and an
angular momentum which prevent the collapse of the knot, and thus
guarantees the dynamical stability of the knot \cite{cho3}.

The Skyrme theory has always been interpreted as an effective
theory of strong interaction, with the skyrmion identified as the
baryon. This interpretation is based on the fact that the Skyrme
theory can be viewed as a non-linear sigma model which describes
the pion physics (in general the flavor dynamics). In this view
the baryon is identified as a topological soliton made of mesons.
If so, one may ask how the knot can be interpreted in this
picture. Since the skyrmion and the knot have different topology
and since the knot has a vanishing baryon number, one may
naturally interpret the knot as a new type of topological
meson. In this view one can estimate the energy of the knot,
which turns out to be around $5~GeV$.

In this paper we provide an alternative view. The alternative view
follows from the observation made by Faddeev and Niemi
that the Skyrme theory is closely related to QCD \cite{fadd2}.
Based on this we argue that the Skyrme theory can be viewed
as an effective theory of strong interaction which is
dual to QCD. This is because QCD is a theory of confinement
in which the chromoelectric flux of the quarks and gluons
are confined by the dual Meissner effect, but the Skyrme theory
can be viewed as a theory of monopole in which
the magnetic flux of monopole-antimonopole pairs are
confined by the Meissner effect \cite{cho01,cho3}.
Obviously this dual picture is completely orthogonal to the
popular view that the Skyrme theory describes the flavor dynamics,
because this view advocates that the Skyrme theory describes the
chromomagnetic dynamics. Nevertheless this alternative
interpretation is worth a serious consideration, and has an
interesting prediction. Based on this dual picture between
Skyrme theory and QCD, we predict the existence of a
chromoelectric knot in QCD which is dual to Faddeev-Niemi knot, a
topological glueball of the twisted chromoelectric flux ring. We
estimate the mass and decay width of such exotic knot glueball to
be around $60~GeV$ and $8~GeV$ \cite{plb05}.

The interpretation of the Faddeev-Niemi knot as a twisted magnetic
vortex ring suggests that it could also be viewed as a topological
object in condensed matter. In fact, recently similar knots have
been asserted to exist in condensed matter physics, in particular
in multi-component Bose-Einstein condensates \cite{cho1,ruo} and
in multi-gap superconductors \cite{cho2,baba}.  In this paper we
point out a remarkable similarity which exists between these knots
in condensed matters and the Faddeev-Niemi knot in Skyrme theory.
In particular we show that the Skyrme-Faddeev theory can also be
viewed as a theory of self-interacting two-component superfluid,
in which the non-linear Skyrme interaction describes the vorticity
interaction of the superfluid. This provides yet another
interpretation of the Faddeev-Niemi knot, two quantized vorticity
flux rings linked together in two-component superfluid. This
strongly indicates that the Skyrme theory, with the built-in
Meissner effect, could also describe a very interesting low energy
condensed matter physics.

The paper is organized as follows. In Section II we review the the
various topological objects and their relationships with singular
monopoles in Skyrme theory for later purpose. In particular we
show that the skyrmion can be viewed as a dressed monopole whose
energy is made finite by the dressing of the massless scalar field
of Skyrme theory. In Section III we discuss the helical baby
skyrmion, a twisted chromomagnetic vortex which is periodic in
$z$-coordinate, in Skyrme theory. As importantly we establish the
Meissner effect which confines the magnetic flux of the baby
skyrmion in Skyrme theory. In Section IV we present a numerical
solution for an axially symmetric Faddeev-Niemi knot, and show
that the knot is nothing but a twisted magnetic vortex ring made
of a helical baby skyrmion. In Section V we estimate the mass of
Faddeev-Niemi knot, and discuss the physical significance of the
knot in Skyrme theory. In Section VI we discuss the deep
connection between Skyrme theory and (massive) $SU(2)$ gauge
theory, and show that the Skyrme theory can describe the
chromomagnetic dynamics of $SU(2)$ QCD. In Section VII we propose
the existence of a chromoelectric knot in QCD, based on the dual
relationship between Skyrme theory and QCD. We estimate the mass
and decay width of the lightest chromoelectric knot. In Section
VIII we argue that, with the built-in Meissner effect, the Skyrme
theory could also describe an interesting condensed matter
physics. In particular, we point out an apparent similarity
between the Skyrme theory and the $U(1)$ gauge theory of
two-component Bose-Einstein condensates and superfluids. Finally
in Section IX we discuss the physical implications of our results.

\section{Skyrme Theory: A Review}

The Skyrme theory has long been interpreted as an effective field
theory of flavor dynamics in strong interaction with a remarkable success
\cite{prep,rho,witt}. However, it can also be interpreted as a
theory of monopole in which the monopole-antimonopole pairs are
confined through the Meissner effect \cite{cho01,cho3}. It has
singular non-Abelian monopoles very similar to the Wu-Yang
monopole in $SU(2)$ QCD which play the key role in the theory. All
finite energy topological objects in the theory appear as dressed
monopoles or confined magnetic flux of monopole-antimonopole pair.
To see this, we review the topological objects in Skyrme theory
and their relations first.

Let $\omega$ and $\hat n$ $({\hat n}^2 = 1)$ be the massless
scalar field and the non-linear sigma field in Skyrme theory, and
let
\bea
&L_\mu = U\partial_\mu U^{\dagger}, \nn\\
&U = \exp (\dfrac{\omega}{2i} \vec \sigma \cdot \hat n) = \cos
\dfrac{\omega}{2} - i (\vec \sigma \cdot \hat n)
\sin \dfrac{\omega}{2}.
\label{su2}
\eea
With this one can write the Skyrme Lagrangian as \cite{skyr}
\bea
&{\cal L} = \dfrac{\mu^2}{4} {\rm tr} ~L_\mu^2 +
\dfrac{\alpha}{32}
{\rm tr} \left( \left[ L_\mu, L_\nu \right] \right)^2 \nn\\
&= - \dfrac{\mu^2}{4} \Big[ \dfrac{1}{2} (\partial_\mu \omega)^2
+2 \sin^2 \dfrac{\omega}{2} (\partial_\mu \hat n)^2 \Big] \nn \\
&-\dfrac{\alpha}{16} \Big[ \sin^2 \dfrac{\omega}{2}  (\partial_\mu
\omega \partial_\nu \hat n
-\partial_\nu \omega \partial_\mu \hat n)^2 +4 \sin^4
\dfrac{\omega}{2} (\partial_\mu \hat n \times \partial_\nu \hat
n)^2 \Big],
\label{slag}
\eea
where $\mu$ and $\alpha$ are the coupling constants. The
Lagrangian has a hidden $U(1)$ gauge symmetry as well as a global
$SU(2)$ symmetry. From the Lagrangian one has the following
equations of motion
\bea
&\partial^2 \omega -\sin\omega (\partial_\mu \hat n)^2
+\dfrac{\alpha}{8 \mu^2} \sin\omega (\partial_\mu \omega
\partial_\nu \hat n -\partial_\nu \omega \partial_\mu \hat n)^2
\nn \\
&+\dfrac{\alpha}{\mu^2} \sin^2 \dfrac{\omega}{2}
\partial_\mu \big[ (\partial_\mu \omega \partial_\nu \hat n
-\partial_\nu \omega \partial_\mu \hat n)
\cdot \partial_\nu \hat n \big]
- \dfrac{\alpha}{\mu^2} \sin^2 \dfrac{\omega}{2}
\sin\omega (\partial_\mu \hat n \times
\partial_\nu \hat n)^2 =0, \nn \\
&\partial_\mu \Big\{\sin^2 \dfrac{\omega}{2}  \hat n \times
\partial_\mu \hat n + \dfrac{\alpha}{4\mu^2} \sin^2 \dfrac{\omega}{2}
\big[ (\partial_\nu \omega)^2 \hat n \times \partial_\mu \hat n
-(\partial_\mu \omega \partial_\nu \omega) \hat n \times
\partial_\nu \hat n \big] \nn\\
&+\dfrac{\alpha}{\mu^2} \sin^4 \dfrac{\omega}{2} (\hat n \cdot
\partial_\mu \hat n \times
\partial_\nu \hat n) \partial_\nu \hat n \Big\}=0.
\label{skeq1}
\eea
Notice that the second equation can be interpreted as the
conservation of $SU(2)$ current originating from the global
$SU(2)$ symmetry of the theory. With the spherically symmetric
ansatz
\bea
\omega = \omega (r),~~~~~\hat n = \pm \hat r,
\label{skans}
\eea
(3) is reduced to
\bea
\dfrac{d^2 \omega}{dr^2} +\dfrac{2}{r} \dfrac{d\omega}{dr}
-\dfrac{2\sin\omega}{r^2} +\dfrac{2\alpha}{\mu^2}
\Big[\dfrac{\sin^2 (\omega/2)}{r^2}
\dfrac{d^2 \omega}{dr^2}
+\dfrac{\sin\omega}{4 r^2} (\dfrac{d\omega}{dr})^2
-\dfrac{\sin\omega \sin^2 (\omega /2)}{r^4} \Big] =0.
\label{skeq2}
\eea
Notice that the energy of the spherically symmetric solutions is
given by
\bea
&E = \dfrac{\pi}{2} \mu^2 \int^{\infty}_{0}
\bigg\{\left(r^2 + \dfrac{2\alpha}{\mu^2}
\sin^2{\dfrac{\omega}{2}}\right)\left(\dfrac{d\omega}{dr}\right)^2
+8 \left(1 + \dfrac{\alpha}{2\mu^2~r^2}\sin^2{\dfrac{\omega}{2}}
\right)
\sin^2 \dfrac{\omega}{2} \bigg\} dr \nn\\
&= \pi {\sqrt \alpha} \mu \dfrac{}{} \int^{\infty}_{0}
\Big[x^2 \left(\dfrac{d\omega}{dx}\right)^2
+ 8 \sin^2{\dfrac{\omega}{2}} \Big] dx,
~~~~~\big(x=\dfrac{\mu}{\sqrt \alpha}~r \big)
\label{sken}
\eea
where $x$ is a dimensionless variable. Notice that the last
equality follows from the virial theorem. Imposing the boundary
condition
\bea
\omega(0)= 2\pi,~~~~~\omega(\infty)= 0,
\label{skbc}
\eea
one has the well-known skyrmion which has a finite energy
\cite{skyr,yang}
\bea
E \simeq 73~{\sqrt \alpha}
\mu.
\label{ske2}
\eea
It carries the baryon number
\bea
&Q_s =
\dfrac{1}{24\pi^2} \int
\epsilon_{ijk} ~{\rm tr} ~(L_i L_j L_k) d^3r \nn\\
&= \dfrac{\mp 1}{8\pi^2} \int \epsilon_{ijk} \partial_i \omega
\big[\hat r \cdot (\partial_j \hat r \times \partial_k \hat r) \big]
\sin^2 \dfrac{\omega}{2} d^3r \nn\\
&=\pm 1,
\label{bn}
\eea
which represents the non-trivial homotopy $\pi_3(S^3)$ defined by
$U$ in (\ref{su2}).

A remarkable point of (\ref{skeq1}) is that
\bea
\omega= \pi,
\eea
becomes a classical solution, independent of $\hn$ \cite{cho01}.
So restricting $\omega$ to $\pi$, one can reduce the Skyrme
Lagrangian (\ref{slag}) to the Skyrme-Faddeev Lagrangian
\bea
{\cal L} \rightarrow -\dfrac{\mu^2}{2} (\partial_\mu \hat
n)^2-\dfrac{\alpha}{4}(\partial_\mu \hat n \times
\partial_\nu \hat n)^2,
\label{sflag}
\eea
whose equation of motion is given by
\bea
&\hn \times \partial^2 \hn + \dfrac{\alpha}{\mu^2} (\partial_\mu
H_{\mu\nu})
\partial_\nu \hn = 0, \nn\\
&H_{\mu\nu} = \hn \cdot (\partial_\mu \hn \times \partial_\nu \hn)
= \partial_\mu C_\nu - \partial_\nu C_\mu.
\label{sfeq}
\eea
Notice that $H_{\mu\nu}$ allows a potential $C_\mu$ because it
forms a closed two-form. Again the equation can be viewed as a
conservation of $SU(2)$ current,
\bea
\partial_\mu (\hn \times \partial_\mu \hn
+\dfrac{\alpha}{\mu^2} H_{\mu\nu}\partial_\nu \hn) = 0. \nn
\eea
It is this equation that allows not only the baby skyrmion and the
Faddeev-Niemi knot but also the non-Abelian monopole.

Just like the $SU(2)$ QCD the Lagrangian (\ref{sflag}) has
singular Wu-Yang type monopole solutions \cite{cho01,cho3}
\bea
\hn = \pm \hat r.
\label{mono}
\eea
They become solutions of (\ref{sfeq}) except at the origin,
because
\bea
&\partial^2 \hat r = - \dfrac {2}{r^2} \hat r, \nn\\
&\partial_\mu \big[\hat r \cdot (\partial_\mu \hat r
\times \partial_\nu \hat r) \big] =0.
\eea
Moreover we have
\bea
H_{\hat r}=\pm \dfrac{\hat r}{r^2}, ~~~~~H_{\hat
\theta}=0,~~~~~H_{\hat \varphi}=0, \nn
\eea
so that it carries the magnetic charge \cite{cho01,cho1}
\bea
&Q_m = \dfrac{\pm 1}{8\pi} \int \epsilon_{ijk} \big[\hat r
\cdot (\partial_i \hat r \times \partial_j \hat r)\big] d\sigma_k \nn\\
&= \pm 1,
\eea
which represents the homotopy $\pi_2(S^2)$ defined by $\hn$. This
is precisely the magnetic field of a singular monopole located at
the origin, which is very similar to the Wu-Yang monopole. In
$SU(2)$ QCD we have the well known Wu-Yang monopole
\cite{cho80,cho82}
\bea
&\vec A_\mu= -\dfrac{1}{g} \hat r \times \partial_\mu \hat r,
&\vec F_{\mu\nu}= -\dfrac{1}{g} \partial_\mu \hat r
\times \partial_\nu \hat r,
\eea
whose magnetic field is defined by
\bea
\tilde H_{\mu\nu}= -\dfrac{1}{g} \hat r \cdot (\partial_\mu \hat r
\times \partial_\nu \hat r).
\eea
This is almost identical to the monopole solution (\ref{mono}),
which justifies us to interpret $C_\mu$ and $H_{\mu\nu}$ as the
magnetic potential and magnetic field generated by the monopole.

However, there are also significant differences between the two
monopoles.First, in QCD $\hat n=+\hat r$ and $\hat n=-\hat r$ are
gauge equivalent, whereas here they are physically different
because the Skyrme theory has no local $SU(2)$ symmetry. So,
unlike in QCD,  the monopole and the anti-monopole in Skyrme
theory are not equivalent. Secondly, our monopole here has the
quadratic interaction in the Lagrangian (\ref{sflag}). Because of
this, the energy of the monopoles has divergent contribution from
both the origin and the infinity. In contrast, the energy of
Wu-Yang monopole in QCD is divergent only at the origin. Finally,
the above solution becomes a solution even without the quartic
interaction (i.e., with $\alpha=0$). This justifies the
interpretation that the Skyrme theory is indeed a theory of
monopole (interacting with the massless scalar field $\omega$).
But one has to keep in mind that this monopole is not an
electromagnetic monopole, but rather a non-Abelian chromomagnetic
one.

Actually the Skyrme theory has more complicated monopole
solutions. To see this consider (\ref{skeq2}) with the spherical
symmetry (\ref{skans}) again, and impose the following boundary
condition
\bea
\omega(0)=\pi,~~~~~\omega(\infty)=0,
\label{+mbc}
\eea
With this one can find a monopole solution which reduces to the
singular solution (\ref{mono}) near the origin. The only
difference between this and the singular solution is the
non-trivial dressing of the scalar field $\omega$, so that it
could be interpreted as a dressed monopole. This dressing,
however, is only partial because this makes the energy finite at
the infinity, but not at the origin. One can also impose the
following boundary condition
\bea
\omega(0)=0,~~~~~\omega(\infty)=-\pi.
\label{-mbc}
\eea
and obtain another monopole solution which approaches the singular
solution near the infinity. Here again the partial dressing makes
the energy finite at the origin, but not at the infinity. So the
partially dressed monopoles still carry an infinite energy.

Obviously the dressed monopoles have a unit monopole charge $Q_m$,
but carry a half baryon number due to the boundary conditions
(\ref{+mbc}) and (\ref{-mbc}). This must be clear from the
definition of the baryon number (\ref{bn}). In this sense they
could be called half-skyrmions.

The physical significance of the partially dressed monopoles is
not that they are physical, but that together they can form a
finite energy soliton, the fully dressed skyrmion. This must be
clear because putting the boundary conditions (\ref{+mbc}) and
(\ref{-mbc}) together one recovers the boundary condition
(\ref{skbc}) of the skyrmion (modulo $2\pi$) whose energy is
finite. From this one can conclude that the skyrmion is a finite
energy monopole in which the cloud of the massless scalar field
$\omega$ regularizes the energy of the singular monopole both at
the origin and the infinity.

\section{Helical Baby Skyrmion}

It has been well-known that the Skyrme theory has a vortex
solution known as the baby skyrmion \cite{piet}. But the theory
also has a twisted vortex solution, the helical baby skyrmion
\cite{cho3}.

To construct the desired helical vortex we choose the cylindrical
coordinates $(\varrho,\varphi,z)$, and adopt the ansatz
\bea
&\n= \Bigg(
\matrix{
\sin{f(\varrho)} \cos{(n\varphi+mkz)}
\cr
\sin{f(\varrho)} \sin{(n\varphi+mkz)} \cr
\cos{f(\varrho)}
}
\Bigg), \nn\\
&C_\mu = \big(\cos{f(\varrho)} + 1 \big) (n\pro_\mu \varphi + m k
\pro_\mu z).
\label{hvans}
\eea
With this the equation (\ref{sfeq}) is reduced to
\bea
&\Big(1+\dfrac{\alpha}{\mu^2}(\dfrac{n^2}{\varrho^2}+m^2 k^2)
\sin^2{f}\Big) \ddot{f} \nn\\&+ \Big( \dfrac{1}{\varrho}
+\dfrac{\alpha}{\mu^2}(\dfrac{n^2}{\varrho^2}+m^2 k^2) \dot{f}
\sin{f}\cos{f} \nn\\
&- \dfrac{\alpha}{\mu^2}\dfrac{1}{\varrho}
(\dfrac{n^2}{\varrho^2}-m^2 k^2) \sin^2{f} \Big) \dot{f} \nn\\
&- (\dfrac{n^2}{\varrho^2}+m^2 k^2) \sin{f}\cos{f}=0.
\label{hveq}
\eea
So with the boundary condition
\bea
f(0)=\pi,~~f(\infty)=0,
\label{bc}
\eea
we obtain the non-Abelian vortex solutions shown in
Fig.~\ref{hbs}. When $m=0$, the solution describes the well-known
baby skyrmion \cite{piet}. But when $m$ is not zero, it describes
a helical vortex which is periodic in $z$-coordinate \cite{cho3}.
Our result shows that the size (radius) of vortex can drastically
be reduced by twisting.

\EPSFIGURE[t]{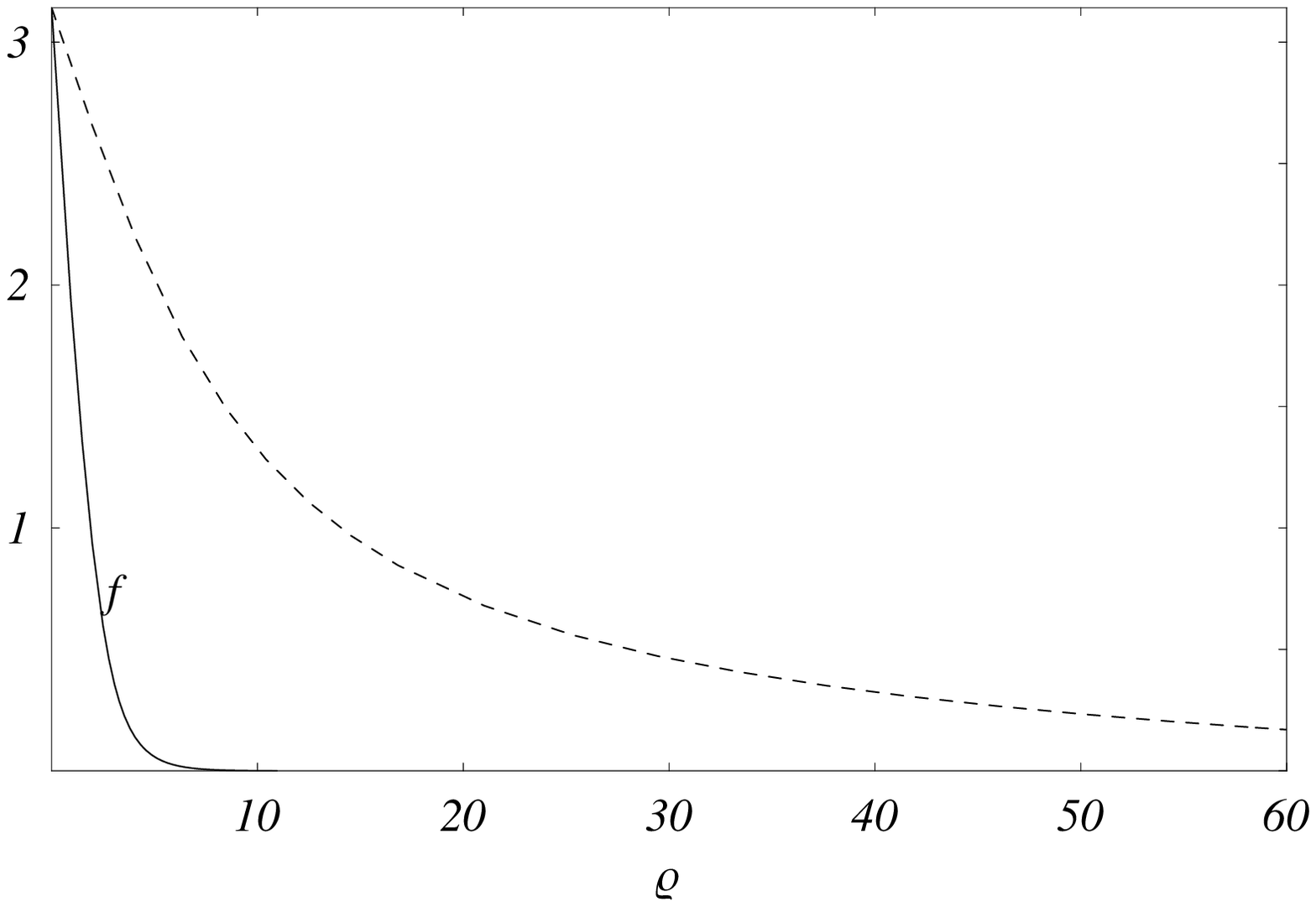,width=9cm}{The baby skyrmion (dashed line) with $m=0,n=1$
and the helical baby skyrmion (solid line) with $m=n=1$ in Skyrme
theory. Here $\varrho$ is in the unit ${\sqrt \alpha}/\mu$ and
$k=0.8~\mu/{\sqrt \alpha}$. \label{hbs}}

Notice that the helical vortex has a non-vanishing magnetic
potential $C_\mu$ (not only around the vortex but also) along the
$z$-axis, so that it has two helical magnetic fields
\bea
&H_{\hat{z}}=\dfrac{1}{\varrho}H_{\varrho\varphi}
=-\dfrac{n}{\varrho}\dot{f}\sin{f}, \nn\\
&H_{\hat{\varphi}}=H_{z \varrho}= mk \dot{f}\sin{f},
\eea
which gives two quantized chromomagnetic fluxes. It has a
quantized magnetic flux along the $z$-axis
\bea
&\Phi_{\hat z} = \dfrac {}{}\int H_{\varrho\varphi} d\varrho
d\varphi = 4\pi n,
\label{nqn}
\eea
and a quantized magnetic flux around the $z$-axis (in one period
section from $z=0$ to $z=2\pi/k$)
\bea
&\Phi_{\hat \varphi} = \dfrac {}{}\int H_{z \varrho} d\varrho dz =
-4\pi m.
\label{mqn}
\eea
This confirms that the magnetic fluxes are quantized in the unit
of $4\pi$, the unit of the monopole.

The origin of this quantization of the magnetic flux, of course,
is topological. To see this consider the baby skyrmion with $m=0$
(the straight vortex) first. In this case $\hn$ defines a mapping
$\pi_2(S^2)$ from the compactified $xy$-plane $S^2$ to the target
space $S^2$. And the quantized magnetic flux $\Phi_{\hat z}$
describes the winding number of this mapping. Similarly
$\Phi_{\hat \varphi}$ describes the winding number of another
mapping $\pi_2(S^2)$ from the compactified half $xz$-plane to the
target space $S^2$.

The vortex solutions implies the existence of Meissner effect
which confines the magnetic flux of the vortex \cite{cho01,cho3}.
To see how the Meissner effect comes about, notice that due to the
$U(1)$ gauge symmetry the Skyrme theory has a conserved current,
\bea
&j_\mu = \pro_\nu H_{\mu\nu},~~~~~\pro_\mu j_\mu = 0.
\eea
So the magnetic flux of the vortex can be thought to come from the
helical chromoelectric supercurrent density
\bea
&j_\mu = -(\pro^2 C_\mu-\pro_\mu \pro_\nu C_\nu)   \nn\\
&=n\varrho \dfrac{d}{d\varrho}
\Big(\dfrac{\sin f} {\varrho} \dot f \Big) \partial_{\mu} \varphi
+mk \dfrac{d}{\varrho d\varrho}
\Big(\varrho \dot f \sin f \Big) \partial_{\mu}z \nn\\
&=\sin f \Big[n \big(\ddot f + \dfrac{\cos f}{\sin f}
\dot f^2 - \dfrac{1}{\varrho} \dot f \big) \partial_{\mu}\varphi \nn\\
&+mk \big(\ddot f + \dfrac{\cos f}{\sin f} \dot f^2 +
\dfrac{1}{\varrho} \dot f \big) \partial_{\mu}z \Big].
\label{cc}
\eea
This produces the supercurrents $i_{\hat\varphi}$ (per one period
section from $z=0$ to $z=2\pi/k$) around the $z$-axis
\bea
&i_{\hat\varphi} = n \dfrac{}{}\int_{\varrho=0}^{\varrho=\infty}
\int_{z=0}^{z=2\pi/k}
\sin f \big(\ddot f + \dfrac{\cos f}{\sin f} \dot f^2 \nn\\
&- \dfrac{1}{\varrho} \dot f \big)  \dfrac{d\varrho}{\varrho} dz
=\dfrac{2 \pi n}{k}\dfrac{\sin{f}}{\varrho}\dot f
\Bigg|_{\varrho=0}^{\varrho=\infty} \nn\\
&=\dfrac{2 \pi n}{k} \dot f^2(0),
\eea
and $i_{\hat z}$ along the $z$-axis
\bea
&i_{\hat z} = mk \dfrac{}{}\int_{\varrho=0}^{\varrho=\infty}
\sin f \big(\ddot f + \dfrac{\cos f}{\sin f} \dot f^2
+ \dfrac{1}{\varrho} \dot f \big) \varrho d\varrho d\varphi \nn\\
&=2 \pi mk \varrho \dot f \sin{f}
\Bigg|_{\varrho=0}^{\varrho=\infty} =0.
\eea
Notice that, even though $i_{\hat z}=0$, it has a non-trivial
current density.

\EPSFIGURE[t]{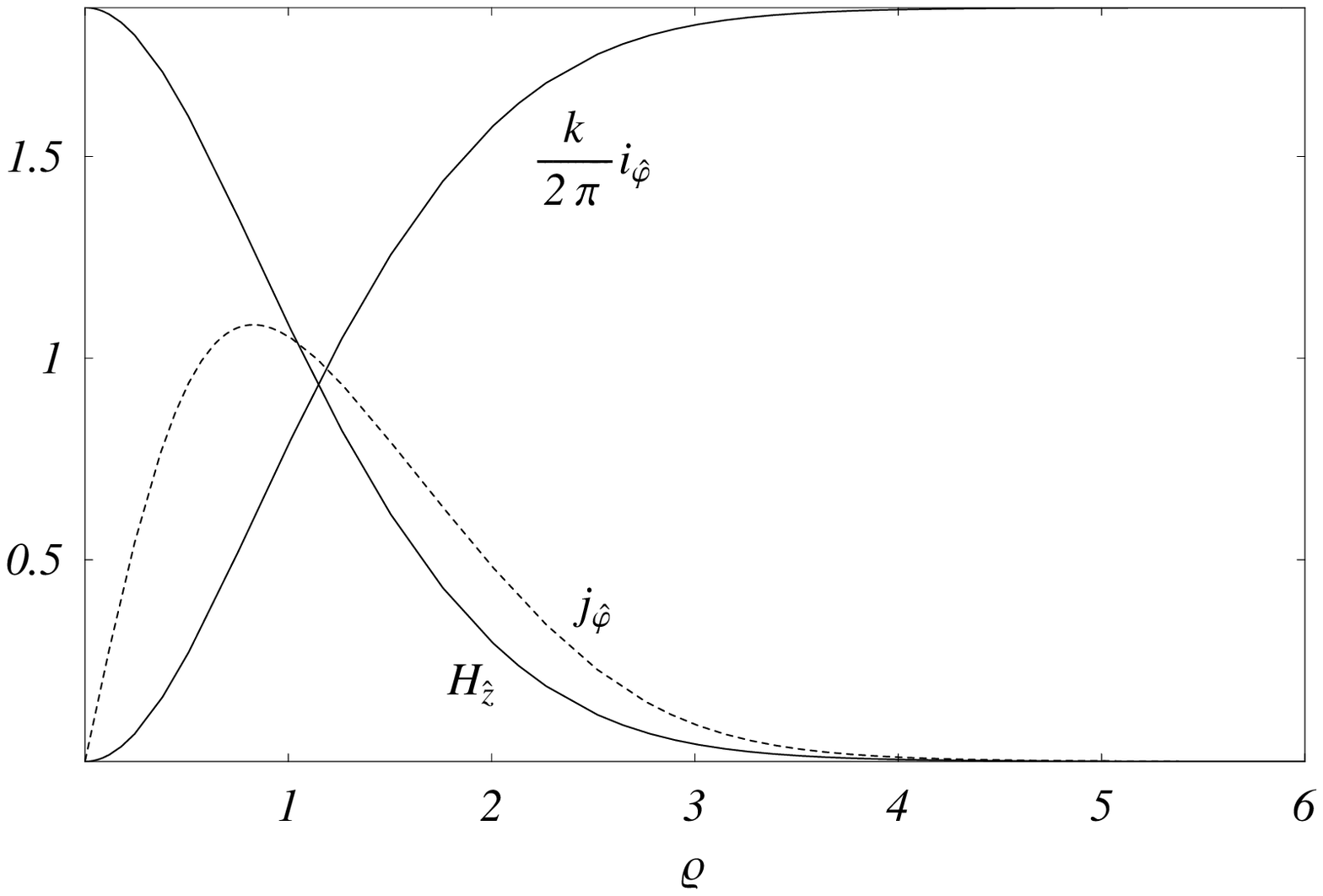, width=9cm}{The supercurrent $i_{\hat \varphi}$ (in one period section
in $z$-coordinate) and corresponding magnetic field $H_{\hat z}$
circulating around the cylinder of radius $\varrho$ of the helical
vortex with $m=n=1$, where $\varrho$ is in the unit ${\sqrt
\alpha}/\mu$ and $k=0.8~\mu/{\sqrt \alpha}$. The current density
$j_{\hat \varphi}$ is represented by the dotted line.
\label{skyiphi}}

The helical magnetic fields and supercurrents are shown in
Fig.~\ref{skyiphi} and Fig.~\ref{skyiz}. Clearly the helical
magnetic fields are confined along the $z$-axis, confined by the
helical supercurrent. This is nothing but the Meissner effect,
which assures that the Skyrme theory has a built-in confinement
mechanism which confines the magnetic flux. Notice that for the
monopole solutions (\ref{mono}) this supercurrent becomes
identically zero, which is why the Meissner effect does not work
for the monopoles.

\EPSFIGURE[t]{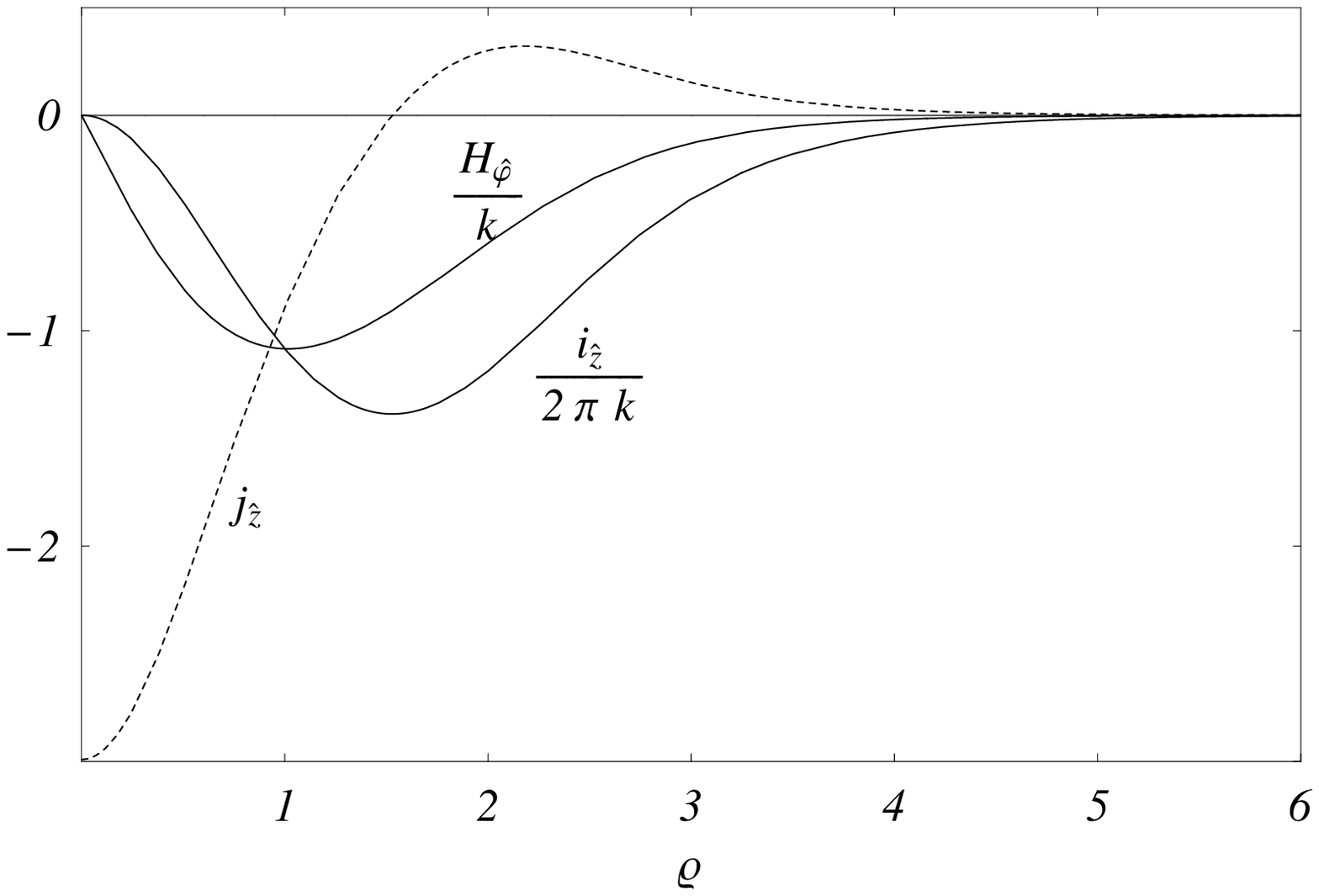, width=9cm}{The supercurrent $i_{\hat z}$ and corresponding
magnetic field $H_{\hat \varphi}$ flowing through the disk of
radius $\varrho$ of the helical vortex with $m=n=1$, where
$\varrho$ is in the unit ${\sqrt \alpha}/\mu$ and
$k=0.8~\mu/{\sqrt \alpha}$. The current density $j_{\hat z}$ is
represented by the dotted line. \label{skyiz}}

With the ansatz (\ref{hvans}) the energy (in one periodic section)
of the helical vortex is given by
\bea
&E = \dfrac{2\pi^2 \mu^2}{k} \dfrac{}{}\int^\infty_0
\bigg\{\left(1 +\dfrac{\alpha}{\mu^2}
(\dfrac{n^2}{\varrho^2}+k^2 m^2)\sin^2{f}\right) {\dot f}^2 \nn \\
&+ \left(\dfrac{n^2}{\varrho^2}+k^2 m^2\right)\sin^2{f}\bigg\}
\varrho d\varrho \nn\\
&= \dfrac{2\pi^2 \mu^2}{k} \dfrac{}{}\int^\infty_0
\bigg\{\left(1 + (\dfrac{n^2}{x^2}
+\dfrac{\alpha}{\mu^2} k^2 m^2)\sin^2{f}\right)\left(
\dfrac{d f}{dx}\right)^2 \nn \\
&+ \left(\dfrac{n^2}{x^2}+\dfrac{\alpha}{\mu^2} k^2 m^2\right)
\sin^2{f}\bigg\} xdx,
\label{hven}
\eea
where
\bea
x=\dfrac{\mu}{\sqrt \alpha} \varrho. \nn
\eea
One could calculate the energy of the helical baby skyrmion
numerically. With $m=n=1$ and $k=0.8~\mu/{\sqrt \alpha}$ we find
\bea
E \simeq 224.5~{\sqrt \alpha} \mu.
\label{hbse}
\eea
In comparison for the baby skyrmion (i.e., for $m=0$ and $n=1$) of
the same length we have
\bea
E \simeq 100.9~{\sqrt \alpha} \mu.
\eea
This shows that twisting requires a considerable amount of energy.

\section{Axially Symmetric Faddeev-Niemi Knot: A Numerical Solution}

The helical vortex in Skyrme theory is unphysical in the sense
that it becomes unstable and decays to the untwisted baby skyrmion
unless the periodicity condition is enforced by hand. But it
allows us to construct a knot \cite{cho01,cho3}. This is because
by smoothly connecting two periodic ends of the helical vortex we
can naturally enforce the periodicity condition and make it a
stable vortex ring. By construction this vortex ring carries two
magnetic fluxes, $m$ unit of flux passing through the disk of the
ring and $n$ unit of flux passing along the ring. Moreover the two
fluxes can be thought of two unit flux rings linked together
winding each other $m$ and $n$ times, whose linking number becomes
$mn$. This implies that the twisted vortex ring becomes a knot.

To confirm this we now construct a knot explicitly with a
consistent ansatz. To do this we first introduce the toroidal
coordinates $(\eta,\gamma,\varphi)$ defined by
\begin{eqnarray}
&x=\dfrac{a}{D}\sinh{\eta}\cos{\varphi},
~~~y=\dfrac{a}{D}\sinh{\eta}\sin{\varphi}, \nn\\
&z=\dfrac{a}{D}\sin{\gamma}, \nn\\
&D=\cosh{\eta}-\cos{\gamma}, \nn\\
&ds^2=\dfrac{a^2}{D^2} \Big(d\eta^2+d\gamma^2+\sinh^2\eta
d\varphi^2 \Big), \nn\\
&d^3x=\dfrac{a^3}{D^3} \sinh{\eta} d\eta d\gamma d\varphi,
\label{tc}
\end{eqnarray}
where $a$ is the radius of the knot defined by $\eta=\infty$. Now
we adopt the following axially symmetric knot ansatz \cite{fadd1},
\bea
&\hn=\Bigg(\matrix{\sin f \cos (n\beta+m\varphi) \cr \sin f
\sin (n\beta + m\varphi) \cr \cos f } \Bigg),
\label{skkans}
\eea
where $f$ and $\beta$ are functions of $\eta$ and $\gamma$. With
the ansatz we have
\begin{eqnarray}
&C_\mu = n(\cos f -1) \partial _\mu \beta
+m(\cos f +1)\partial _\mu \varphi, \nn\\
&H_{\eta \gamma}=-nK \sin f,
~~~~~H_{\gamma \varphi }=-m\sin f\partial_\gamma f, \nn\\
&H_{\varphi \eta }=m \sin f\partial_\eta f, \nn\\
&K = \partial _\eta f\partial _\gamma \beta -\partial _\gamma f
\partial_\eta \beta.
\label{kmf1}
\end{eqnarray}
Notice that, in the orthonormal frame $(\hat \eta, \hat \gamma,
\hat \varphi)$, we have
\begin{eqnarray}
&C_{\hat{\eta}}=\dfrac{nD}{a}(\cos f-1)\partial _\eta \beta,
~~~C_{\hat{\gamma}}=\dfrac {nD}{a}(\cos f-1)\partial_\gamma \beta, \nn\\
&C_{\hat{\varphi}}=\dfrac{mD}{a\sinh \eta}(\cos f+1), \nn\\
&H_{\hat{\eta}\hat{\gamma}}=-\dfrac{nD^2}{a^2} K \sin f,
~~~H_{\hat{\gamma}\hat{\varphi}}=-\dfrac{mD^2}{a^2\sinh \eta}
\sin f\partial_\gamma f, \nn\\
&H_{\hat{\varphi}\hat{\eta}}=\dfrac{mD^2}{a^2\sinh \eta}
\sin f\partial_\eta f.
\label{kmf2}
\end{eqnarray}
With this the knot equation (\ref{sfeq}) is written as
\bea
&\Big[\partial_\eta^2 +\partial_\gamma^2 +\Big(\dfrac{\cosh
\eta}{\sinh \eta} -\dfrac{\sinh \eta}D\Big)\partial_\eta
-\dfrac{\sin \gamma}D\partial_\gamma \Big]f \nn \\
&-\Big(n^2
\big((\partial_\eta \omega)^2 +(\partial_\gamma \omega)^2 \big)
+\dfrac{m^2}{\sinh ^2\eta}\Big) \sin f\cos f  \nn \\
&+\dfrac{\alpha}{\mu^2} \dfrac{D^2}{a^2}
\Big(A\cos f +B\sin f \Big)\sin f =0, \nn\\
&\Big[\partial_\eta^2 +\partial_\gamma^2 +\Big(\dfrac{\cosh
\eta}{\sinh \eta}-\dfrac{\sinh \eta}D \Big)
\partial _\eta -\dfrac{\sin \gamma}D\partial_\gamma \Big]\beta \nn
\\
&+2\Big(\partial_\eta f\partial_\eta \beta +\partial_\gamma
f\partial_\gamma \beta \Big)\dfrac{\cos f}{\sin f}
-\dfrac{\alpha}{\mu^2} \dfrac{D^2}{a^2} C=0,
\label{skeq4}
\eea
where
\begin{eqnarray}
&A=\Big[n^2 K^2+\dfrac{m^2}{\sinh ^2\eta}
\Big((\partial_\eta f)^2+(\partial _\gamma f)^2\Big)\Big], \nn\\
&B = \Big\{n^2 \partial_\eta K\partial_\gamma \beta -n^2
\partial_\gamma K\partial_\eta \beta +n^2 K\Big[\Big(\dfrac{\cosh
\eta}{\sinh \eta} +\dfrac{\sinh \eta }D \Big)\partial_\gamma
\beta-\dfrac{\sin \gamma}D
\partial_\eta \beta \Big] \nn\\
&+\dfrac{m^2}{\sinh^2\eta}\Big[\partial_\eta^2 +\partial_\gamma^2
-\Big(\dfrac{\cosh \eta}{\sinh \eta} -\dfrac{\sinh \eta}D\Big)
\partial_\eta +\dfrac{\sin \gamma}D
\partial_\gamma \Big]f \Big\}, \nn\\
&C=\Big\{\partial_\eta K\partial_\gamma f -\partial_\eta f
\partial_\gamma K +K\Big[\Big(\dfrac{\cosh \eta}{\sinh
\eta}+\dfrac{\sinh \eta}D \Big)
\partial_\gamma -\dfrac{\sin \gamma}D\partial_\eta \Big)f\Big\}. \nn
\end{eqnarray}
From the ansatz (\ref{skkans}) we have the following Hamiltonian
\begin{eqnarray}
&{\cal H}=\dfrac{\mu^2}{2} \dfrac{D^2}{a^2}
\Big[(\partial_\eta f)^2+(\partial_\gamma f)^2
+\Big(n^2 \big((\partial _\eta \beta)^2 +(\partial_\gamma \beta)^2
\big) +\dfrac{m^2}{\sinh^2 \eta}\Big) \sin^2 f \Big] \nn \\
&+\dfrac{\alpha}{4} \dfrac{D^4}{a^4} A \sin^2 f,
\label{skh}
\end{eqnarray}
and the energy of the knot
\begin{eqnarray}
&E=\dfrac{}{} \int {\cal H} \dfrac{a^3}{D^3}
\sinh \eta d\eta d\gamma d\varphi  \nn\\
&= {\sqrt \alpha} \mu \dfrac{}{} \int \Big\{\dfrac{\mu}{\sqrt
\alpha}
\dfrac{a}{2D}\Big[(\partial_\eta f)^2+(\partial_\gamma f)^2
+ \Big(n^2 \big((\partial_\eta \beta)^2 +(\partial_\gamma \beta)^2
\big)
+\dfrac{m^2}{\sinh^2 \eta}\Big) \sin^2 f \Big] \nn\\
&+\dfrac{\sqrt \alpha}{\mu}\dfrac{D}{4a}\Big[n^2 K^2
+\dfrac{m^2}{\sinh ^2\eta} \Big((\partial_\eta f)^2 +(\partial
_\gamma f)^2\Big)\Big] \sin^2 f \Big\}
\sinh \eta d\eta d\gamma d\varphi.
\label{ske}
\end{eqnarray}
Minimizing the energy we reproduce the knot equation
(\ref{skeq4}), which tells that our ansatz (\ref{skkans}) is
indeed consistent.

In toroidal coordinates, $\eta=\gamma=0$  represents spatial
infinity and $\eta=\infty$ describes the torus center. So we can
impose the following boundary condition
\begin{eqnarray}
&f(0,\gamma)=0,
~~~~~f(\infty,\gamma)=\pi, \nn\\
&\beta(\eta,0)=0, ~~~~~\beta(\eta,2 \pi)=2 \pi,
\label{knotbc}
\end{eqnarray}
to obtain the desired knot. Of course, an exact solution of
(\ref{skeq4}) is extremely difficult to obtain, even numerically.
In fact many known ``knot solutions" are actually the energy
profile of knots which minimizes the Hamiltonian
\cite{fadd1,batt}. But here we can find an actual profile of the
knot because we have the explicit knot ansatz (\ref{skkans}). For
$m=n=1$ we find that the knot radius $a$ which minimizes the
energy is given by
\bea
a \simeq 1.21~\dfrac{\sqrt \alpha}{\mu}.
\label{skkrad}
\eea
With this we obtain the knot profile for $f$ and $\omega$ of the
lightest axially symmetric knot shown in Fig.~\ref{skkf} and
Fig.~\ref{skkdo}. The numerical result indicates that the radius
of the vortex (the thickness of the knot) $r_0$ is roughly about
\bea
r_0 \simeq a~{\rm csch}~2 \simeq
\dfrac{1}{3} \dfrac{\sqrt \alpha}{\mu},
\label{vrad}
\eea
which means that the radius of the vortex ring is about 3.6 times
the radius of the vortex. From this we can construct a three
dimensional energy profile of the knot shown in Fig.~\ref{skk3d}.

\EPSFIGURE[t]{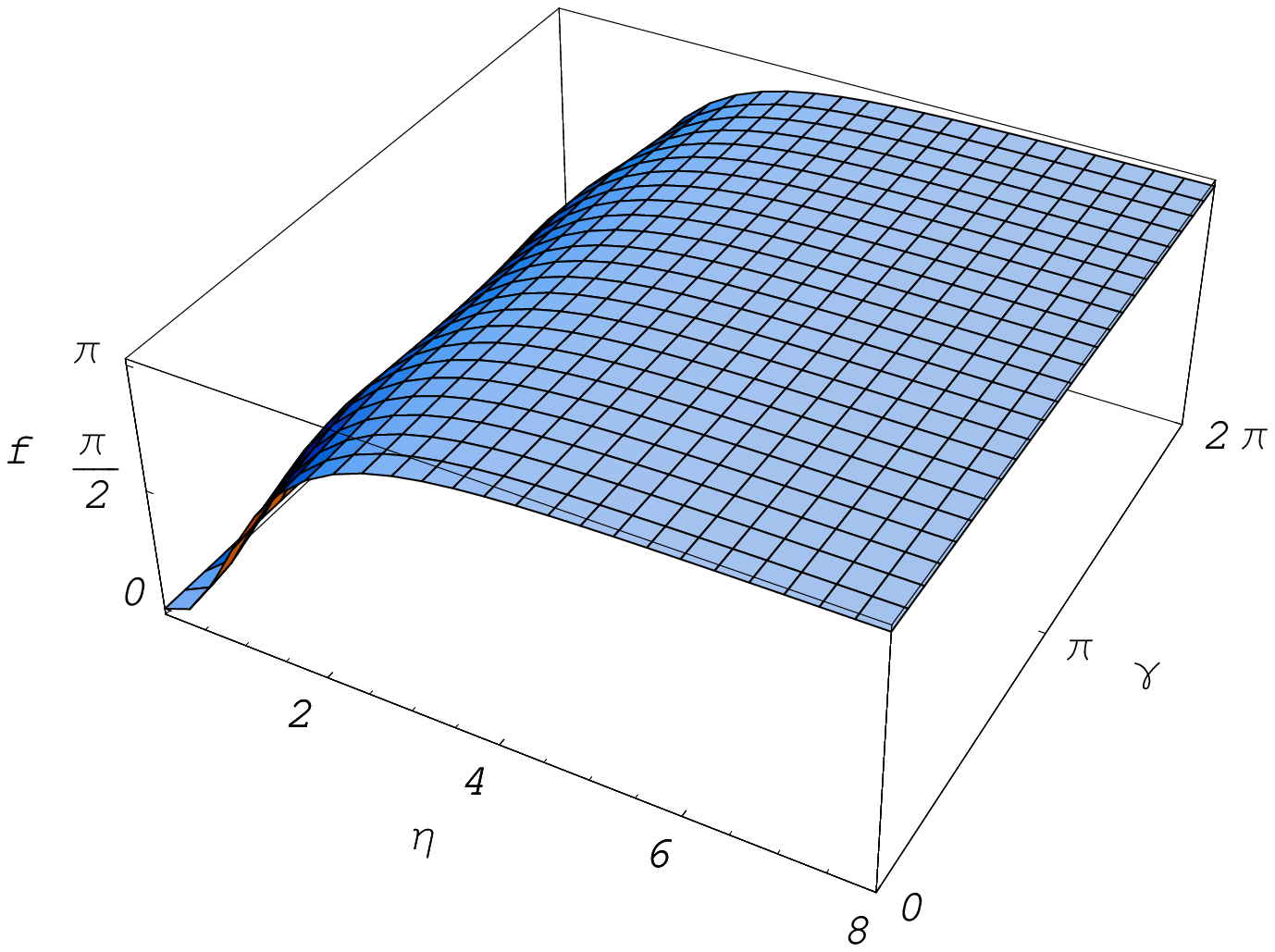, width=9cm}{(Color online). The $f$ configuration of the knot with
$m=n=1$ in Skyrme theory. Notice that here $\eta$ is
dimensionless.
\label{skkf}}

We emphasize that the knot ansatz (\ref{skkans}) has played a
crucial role for us to obtain the numerical solution. With the
ansatz we were able to obtain an actual knot profile, and estimate
the radius of the vortex and the radius of the vortex ring. Notice
that, just for the sake of the knot toplogy, one might have
assumed $\beta=\gamma$ \cite{ward,hiet}. But one can check that
this is inconsistent with the equation of motion, even though this
describes the correct knot topology.

With the numerical solution we can check the topology of the knot.
From the ansatz (\ref{skkans}) we have the knot quantum number
\bea
&Q_k=\dfrac{1}{32\pi^2}\int \epsilon_{ijk}~C_i~H_{jk} d^3x \nn\\
&=\dfrac{mn}{8\pi ^2}\int K \sin f d\eta d\gamma d\varphi
= \dfrac{mn}{4\pi} \int \sin f df d\beta \nn\\
&= mn,
\label{kqn}
\eea
where the last equality comes from the boundary condition
(\ref{knotbc}). This assures that our ansatz describes the correct
knot topology. To understand the meaning of (\ref{kqn}) we now
calculate the magnetic flux of the knot. Since the magnetic field
is helical, we have two magnetic fluxes, $\Phi_{\hat \gamma}$
passing through the knot disk of radius $a$ in the $xy$-plane and
$\Phi_{\hat \varphi}$ passing along the knot ring of radius $a$.
From (\ref{kmf2}) and (\ref{knotbc}) we have
\begin{eqnarray}
&\Phi_{\hat{\gamma}} = \dfrac{}{} \int_{\gamma=\pi}
H_{\hat{\gamma}}
\dfrac{a^2\sinh \eta}{D^2}d\eta d\varphi  \nn\\
&=m\dfrac{}{}\int_{\gamma=\pi} \sin f\partial_\eta f d\eta
d\varphi
=4\pi m, \nn\\
&\Phi_{\hat{\varphi}} = \dfrac{}{} \int H_{\hat{\varphi}}
\dfrac{a^2}{D^2}d\eta d\gamma  \nn\\
&=n \dfrac{}{} \int K \sin fd\eta d\gamma =4\pi n.
\label{kflux}
\end{eqnarray}
This confirms that the flux is quantized in the unit of $4\pi$. As
importantly this tells that the two fluxes are linked, whose
linking number is given by $mn$. This is precisely the knot
quantum number (\ref{kqn}). This proves that the knot quantum
number is given by the linking number of two magnetic fluxes
$\Phi_{\hat \gamma}$ and $\Phi_{\hat \varphi}$. Notice that all
topological quantum numbers are completely fixed by the boundary
condition (\ref{knotbc}).

\EPSFIGURE[t]{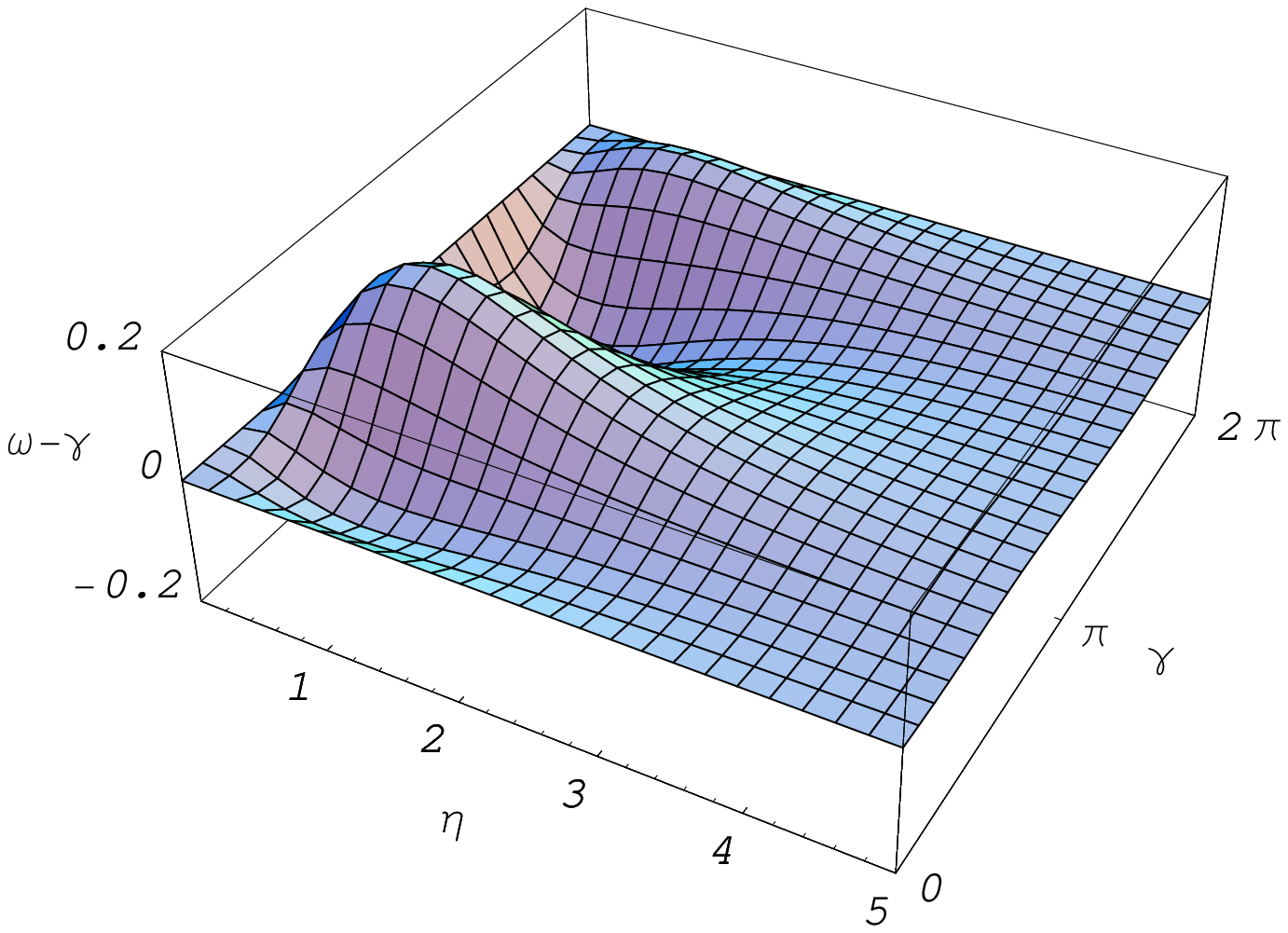, width=9cm}{(Color online). The $\beta$ configuration of the knot
with $m=n=1$ in Skyrme theory. Notice that here the actual
configuration shown is $\beta-\gamma$.
\label{skkdo}}

The supercurrent which generates the twisted magnetic flux of the
knot is given by the conserved current density
\begin{eqnarray}
&j_\mu =\dfrac{nD^2}{a^2}\Big(\partial_\eta +\frac{\cosh \eta}
{\sinh \eta}+\frac{\sinh \eta}D\Big) K \partial_\mu \gamma \nn\\
&+\dfrac{mD^2}{a^2}\Big[\Big(\partial_\eta -\frac{\cosh \eta
}{\sinh \eta}
+\frac{\sinh \eta }D\Big)\sin f\partial_\eta f \nn\\
&+\Big(\partial_\gamma +\dfrac{\sin \gamma}D\Big)\sin f
\partial_\gamma f\Big] \partial_\mu \varphi, \nn\\
&g^{\mu\nu}\nabla_\mu j_\nu =0.
\label{kcd}
\end{eqnarray}
From this we have two quantized supercurrents, $i_{\hat\gamma}$
which flows through the knot disk and $i_{\hat\varphi}$ which
flows along the knot,
\begin{eqnarray}
&i_{\hat{\gamma}} =\dfrac{}{}\int_{\gamma =\pi}
j_{\hat{\gamma}}(\eta,\gamma)
\dfrac{a^2\sinh \eta}{D^2}d\eta d\varphi  \nn\\
&=\dfrac n{a}\int_{\gamma =\pi} D\sinh \eta \Big(\partial_\eta
+\dfrac{\cosh \eta}{\sinh \eta}+\dfrac{\sinh \eta}D\Big)
K d\eta d\varphi,  \nn\\
&i_{\hat{\varphi}}
=\dfrac{}{}\int j_{\hat{\varphi}}\frac{a^2}{D^2}d\eta d\gamma  \nn\\
&=\dfrac m{a}\int \dfrac D{\sinh \eta}
\Big[\Big(\partial_\eta -\dfrac{\cosh \eta}
{\sinh \eta} +\dfrac{\sinh \eta}D\Big)\sin f\partial_\eta f \nn\\
&+\Big(\partial_\gamma +\dfrac{\sin \gamma}D \Big)
\sin f\partial_\gamma f \Big] d\eta d\gamma.
\end{eqnarray}
For $m=n=1$ we find numerically
\bea
&i_{\hat{\gamma}} \simeq \dfrac{23.9}{a}, ~~~~~i_{\hat{\varphi}}
\simeq 0.
\eea
Notice that $i_{\hat{\varphi}}$ is vanishing, which is consistent
with our picture that the knot is a twisted vortex ring made of
the helical vortex. Notice, however, that the current density
$j_{\hat{\varphi}}$ is non-trivial. This tells that the knot has a
net angular momentum around the symmetric axis which stablizes the
knot.

\EPSFIGURE[t]{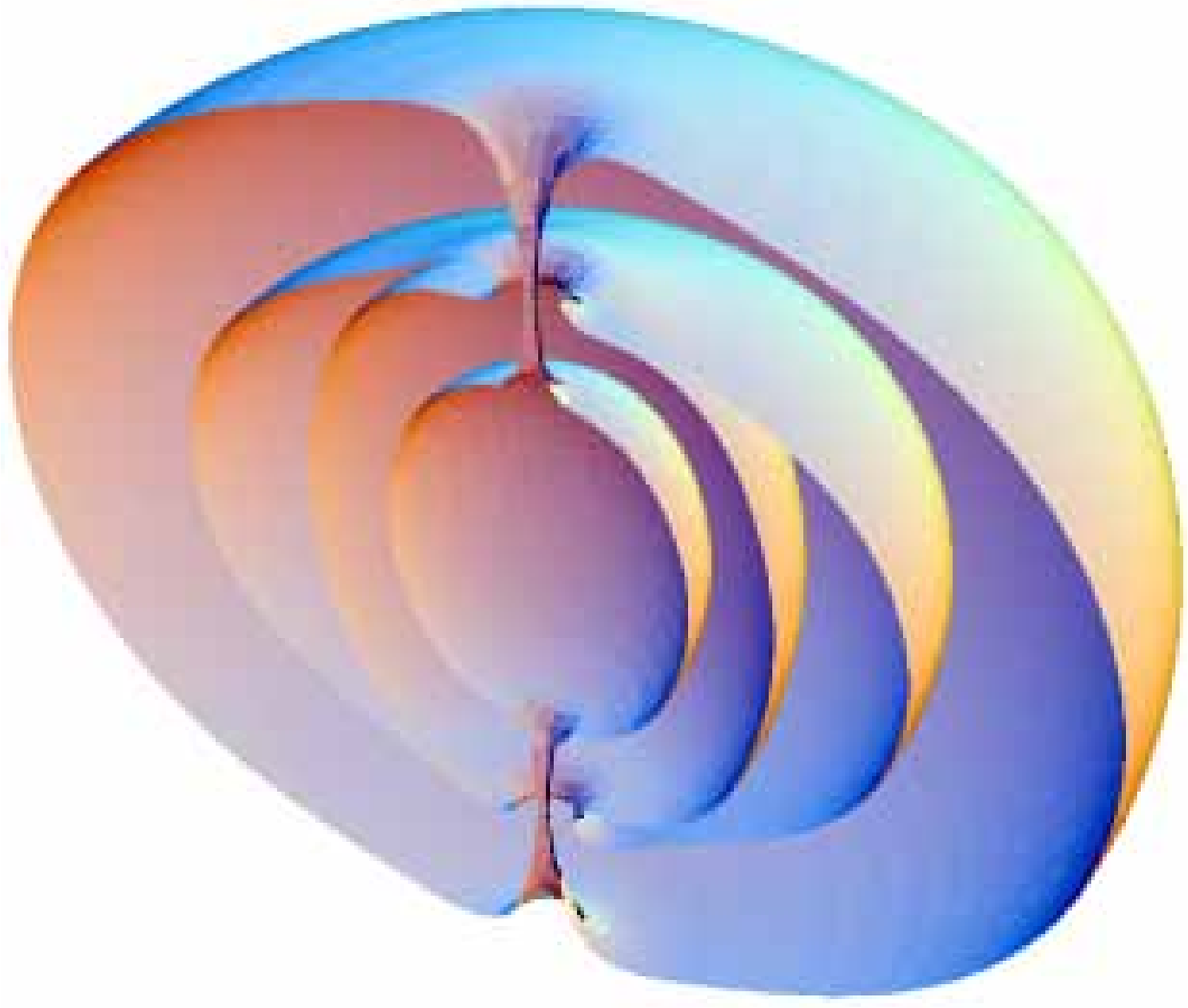,width=9cm}{(Color online). The 3-dimensional energy profile of the lightest
axially symmetric knot with $m=n=1$ in Skyrme theory. Here the
scale is in the unit of ${\sqrt \alpha}/\mu$.
\label{skk3d}}

Our result confirms that the knot solution has a dynamical
manifestation of knot topology \cite{cho3}. In mathematics the
knot topology has always been described by the linking number of
the preimage of the Hopf mapping. Indeed the knot topology of the
Faddeev-Niemi knot is described by the non-linear sigma field
$\hat n$ in (\ref{sfeq}), which defines the Hopf mapping from the
compactified space $S^3$ to the target space $S^2$. When the
preimages of two points of the target space are linked, the
mapping defines a knot. In this case the knot quantum number of
$\pi_3(S^2)$ is given by the linking number of two preimages fixed
by the Chern-Simon index (\ref{kqn}) of the potential $C_\mu$
\cite{fadd1,batt}.

Our interpretation of the Faddeev-Niemi knot as a helical vortex
ring, however, provides an alternative picture of knot. It tells
that the knot is made of two real (physical) magnetic flux rings
linked together, whose knot quantum number is given by the linking
number of two flux rings. This certainly is different from the
above mathematical description of knot based on the Hopf mapping.
This is a dynamical manifestation of knot, the linking of two
physical flux rings which comes from dynamics \cite{cho3}.

Obviously two flux rings linked together can not be unlinked by
any continuous deformation of the field configuration. This
guarantees the topological stability of the knot. Furthermore the
topological stability is backed up by the dynamical stability
which comes from the dynamical manifestation of the knot. This is
because the supercurrent which generates the quantized
chromomagnetic flux of the knot has two components, the component
moving along the knot, and the one moving around the knot tube.
And the supercurrent moving along the knot generates an angular
momentum around the $z$-axis, which provides the centrifugal force
preventing the vortex ring to collapse. Another way to understand
this is to notice that the supercurrent generates the $m$ unit of
the magnetic flux trapped in the knot disk which can not be
squeezed out. Clearly, this flux provides a stabilizing repulsive
force which prevent the collapse of the knot. This is how the knot
acquires the dynamical stability \cite{cho3}.

Our analysis tells that the Faddeev-Niemi knot could also be
viewed as two flux rings made of non-helical baby skyrmion linked
together. Of course, a vortex ring made of straight baby skyrmion
is unstable, because such a ring has no knot topology. But if we
make two such rings and link them together, it becomes a
Faddeev-Niemi knot. Now, two such rings could reconnect and make
one ring. However, in this process the knot topology survives, and
the reconnected ring becomes a helical vortex ring. This tells
that there are actually two complementary ways to view the
Faddeev-Niemi knot, as a vortex ring made of the helical baby
skyrmion or two vortex rings made of the untwisted baby skyrmion
linked together.

The energy of the Faddeev-Niemi knot has been calculated before.
Theoretically it has been known that the energy has the following
bound \cite{ussr},
\bea
c~Q^{3/4} \leq E \leq C~Q^{3/4}.
\eea
Numerically one finds up to $Q_k=8$ \cite{batt}
\bea
E_{Q_k}
\simeq 234~{\sqrt \alpha} \mu~Q^{3/4}.
\label{qke}
\eea
With our ansatz we can estimate the energy of the axially
symmetric knot numerically. For the lightest knot (with $m=n=1$)
we find
\bea
E_1
\simeq 274.0~{\sqrt \alpha}\mu.
\label{skke}
\eea
In general the energy of the axially symmetric knots depends on
(not just $Q$ but) $m$ and $n$, because the axially symmetric knot
ansatz (\ref{skkans}) explicitly depends on them. We have
calculated the energy up to $Q=6$ and obtain the result shown in
Fig.~\ref{skqe}, where we have included the earlier estimate
(\ref{qke}) for comparison (Notice that there are other estimates
of energy based on an inconsistent knot ansatz, which nevertheless
turn out to be similar to our result numerically \cite{hiet}). Our
result shows that the energy of the axially symmetric knots is
systematically larger than the earlier estimate. We do not know
the precise reason for this discrepancy, but this is probably
because of the axial symmetry of the knots. Notice that in reality
there is no reason that the minimum energy knots should have the
axial symmetry, especially for those with $Q_k$ larger than one
\cite{batt}. This implies that the minimum energy knots in general
could have smaller energy than the axially symmetric knots.

\EPSFIGURE[t]{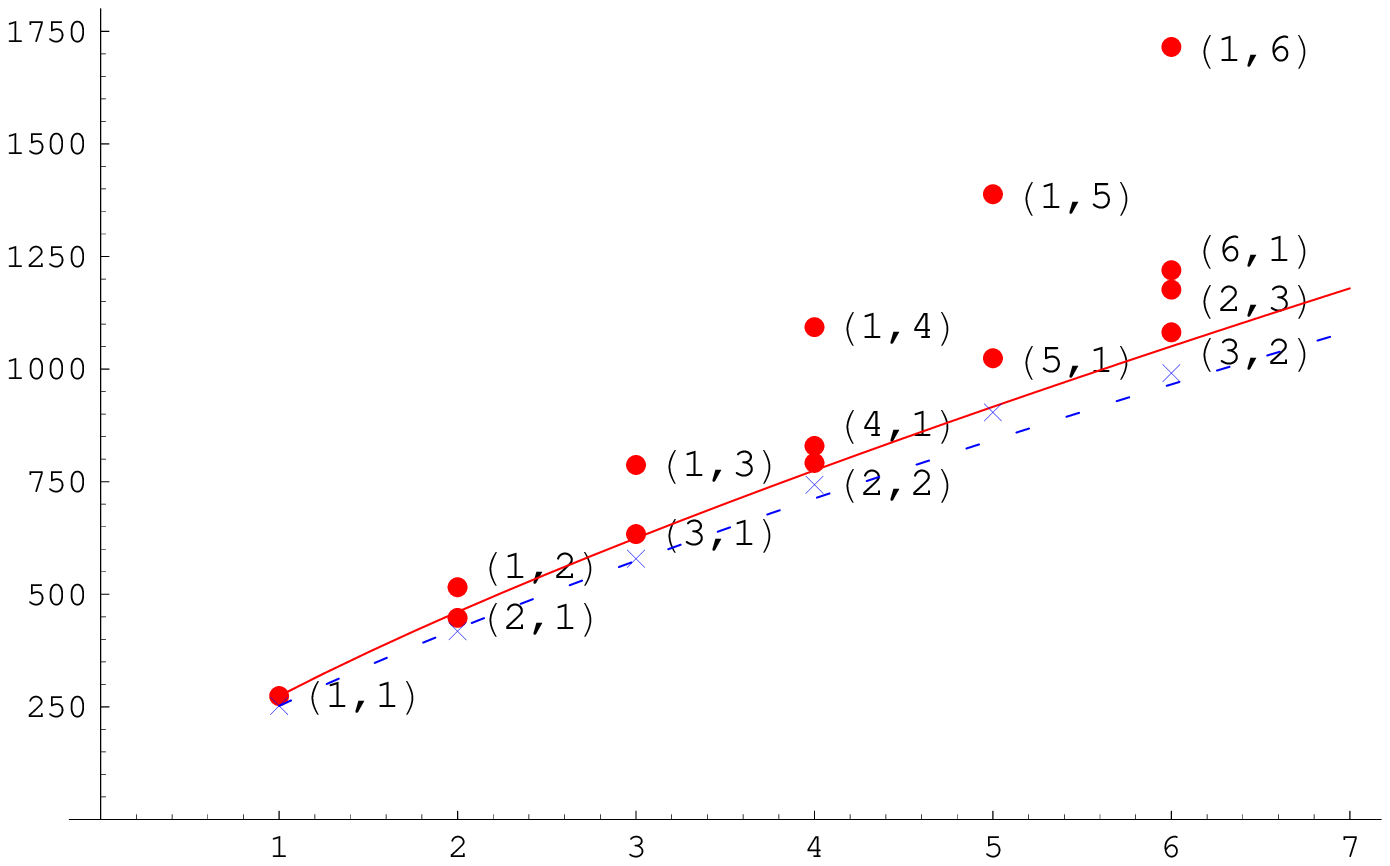, width=9cm}{(Color online). The $Q$-dependence of the energy
of the axially symmetric knots. Our estimate is denoted by dots
and $(m,n)$, with $Q=mn$. For comparison we included the earlier
estimate denoted by crosses, which depends only on $Q$. The solid
and dashed lines represent $E=E_1~Q^{3/4}$ curves.
\label{skqe}}

\section{Physical Interpretation of Knot in Skyrme Theory}

It is well-known that the Skyrme theory can be viewed as a
non-linear sigma model which describes the pion physics
\cite{prep,rho}. Indeed with
\bea
&U = \exp (\dfrac{\omega}{2i} \vec \sigma \cdot \hat n)
=f - i \vec \sigma \cdot \vec \pi, \nn\\
&f=\cos \dfrac{\omega}{2},
~~~~~\vec \pi= \hn \sin \dfrac{\omega}{2}, \nn\\
&f^2 + \vec \pi^2 = 1,
\label{sm}
\eea
the Skyrme Lagrangian (\ref{slag}) can be put into the form
\bea
&{\cal L} = -\dfrac{\mu^2}{2} \big((\partial_\mu f)^2
+(\partial_\mu \vec \pi)^2 \big) \nn\\
&-\dfrac{\alpha}{4} \big((\partial_\mu f \partial_\nu \vec \pi -
\partial_\nu f \partial_\mu \vec \pi)^2
+ (\partial_\mu \vec \pi \times \partial_\nu \vec \pi)^2 \big) \nn\\
&+\dfrac{\lambda}{4} (f^2 + \vec \pi^2 - 1),
\label{smlag}
\eea
where $\lambda$ is a Lagrange multiplier. In this form $f$ and
$\vec \pi$ represent the sigma field and the pion field, so that
the Skyrme theory describes the pion physics. In this picture the
skyrmion becomes a topological soliton made of mesons, which can
be identified as the baryon.

In this picture one may choose the parameters $\mu$ and $\alpha$
to be \cite{prep,rho}
\bea
&\mu \simeq 93~MeV,
~~~~~\alpha \simeq 0.0442, \nn\\
&{\sqrt \alpha}\mu \simeq 19.55~MeV,
\label{data1}
\eea
so that from (\ref{sken}) one has the baryon mass
\bea
m_b \simeq 73~{\sqrt \alpha}\mu \simeq 1.427~GeV.
\label{mb1}
\eea
In a slightly different fitting one may choose \cite{prep,witt}
\bea
&\mu \simeq 65~MeV,
~~~~~\alpha \simeq 0.0336, \nn\\
&{\sqrt \alpha}\mu \simeq 11.91~MeV.
\label{data2}
\eea
to have the baryon mass
\bea
m_b \simeq 0.869~GeV.
\label{mb2}
\eea
In this view one might hope to have composite ``mesons'' which can
be viewed as baryon and anti-baryon bound states. But so far such
bound states have not been found in Skyrme theory, probably
because such states may not exist as topologically stable
solitonic objects. Instead we have the Faddeev-Niemi knot
\cite{cho01}.

Clearly the Faddeev-Niemi knot should describe a non-baryonic
state. Indeed, if the skyrmion is identified as the baryon, the
Faddeev-Niemi knot should be interpreted as a topological meson
(a ``glueball'') made of the twisted flux of pion pair.
In this view one can easily estimate
the mass of the knot from (\ref{qke}), because we know $\mu$
and $\alpha$. With (\ref{data1}) we find the mass of the lightest
knot to be
\bea
m_k \simeq 305.8~{\sqrt \alpha} \mu \simeq 5.98~GeV,
\eea
but with (\ref{data2}) we obtain
\bea
m_k \simeq 3.64~GeV.
\eea
Notice that the above estimate suggests that the knot mass is a
little too big to be viewed as a baryon-antibaryon bound state. As
we have remarked, the knot is not an ordinary bound state, but a
topological meson made of pion pair.

As we have argued, the knot must be stable within the framework of
Skyrme-Faddeev theory. However, one has to be cautious about the
knot stability in Skyrme theory, because here one has an extra
massless scalar field $\omega$ which could (in principle)
destabilize the knot. To see this remember that the knot is made
of the twisted magnetic flux of baryon-antibaryon pair. In
Skyrme-Faddeev theory this flux ring can not be cut, because there
are no finite energy baryon and anti-baryon in the theory which
can terminate the flux at end points. But in Skyrme theory there
are. This suggests that in Skyrme theory, there is a possibility
that the knot could decay to a baryon-antibaryon pair. This
indicates that in Skyrme theory the proof of the stability of the
knot is a non-trivial matter. One needs a careful analysis to
establish the stability of the knot.

It is remarkable that the Skyrme theory allows such diverse
topological objects. In fact it allows all topological objects
known in physics. But what is really remarkable is that the Skyrme
theory itself can be put in many different forms, which put the
theory in a completely different perspective. In the followings we
discuss new interpretations of Skyrme theory
\cite{cho3,cho1,cho2}.

\section{Skyrme theory and QCD}

The Skyrme theory is known to be closely related to QCD \cite{fadd2}.
In this section we expand this fact and show that
the Skyrme theory can be viewed
as an effective theory of strong interaction which is dual to
$SU(2)$ QCD \cite{cho01,cho3}. To show this notice that the
Skyrme-Faddeev Lagrangian (\ref{sflag}) can be put into a very
suggestive form \cite{cho01,cho02},
\bea
&{\cal L} = -\dfrac{\alpha}{4} g^2 \hat H_{\mu\nu}^2
- \dfrac{\mu^2} {2} g^2 \hat C_\mu^2, \nn\\
&\hat H_{\mu\nu} = \partial_\mu \hat C_\nu - \partial_\nu \hat
C_\mu + g \hat C_\mu \times \hat C_\nu,
\label{qcdlag}
\eea
where $\hat C_\mu$ is the well-known magnetic potential of $SU(2)$
QCD \cite{cho80,cho81,fadd2,gies}
\bea
\hat C_\mu = -\dfrac{1}{g} \hn \times \partial_\mu \hn.
\label{ccon}
\eea
This shows that formally the Skyrme-Faddeev Lagrangian
(\ref{sflag}) can be viewed as a massive $SU(2)$ QCD. In general,
the Skyrme Lagrangian (\ref{slag}) itself can be expressed as
\bea
&{\cal L} = - \dfrac{\mu^2}{4} \Big[ \dfrac{1}{2} (\partial_\mu
\omega)^2
+2 g^2 \sin^2 \dfrac{\omega}{2} \hat C_\mu^2 \Big] \nn \\
&- \dfrac{\alpha}{16} g^2 \Big[ \sin^2 \dfrac{\omega}{2}
(\partial_\mu
\omega \hat C_\nu - \partial_\nu \omega \hat C_\mu)^2
+4 \sin^4 \dfrac{\omega}{2} \hat H_{\mu\nu}^2 \Big] \nn\\
&=-\dfrac{\alpha}{4} g^2 (1-\sigma^2)^2 \hat H_{\mu\nu}^2
-\dfrac{\mu^2}{2} g^2 (1-\sigma^2) \hat C_\mu^2 \nn\\
&-\dfrac{\mu^2}{2} \dfrac{(\partial_\mu \sigma)^2}{1-\sigma^2}
-\dfrac{\alpha}{4} g^2 (\partial_\mu
\sigma \hat C_\nu - \partial_\nu \sigma \hat C_\mu)^2,
\label{slag1}
\eea
where
\bea
\sigma=\cos~\dfrac{\omega}{2}. \nn
\eea
So the Skyrme theory could be expressed (with the scalar field
$\omega$) in terms of the magnetic potential (\ref{ccon}) of
$SU(2)$ QCD.

Notice that for small $\sigma$ we can approximate the Lagrangian
(\ref{slag1}) near $\sigma \simeq 0$ with a linear approximation.
Neglecting the higher order interactions we have the linearized
Skyrme Lagrangian
\bea
&{\cal L} \simeq -\dfrac{\alpha}{4} g^2 \hat H_{\mu\nu}^2
-\dfrac{\mu^2}{2} g^2  \hat C_\mu^2 \nn\\
&-\dfrac{\mu^2}{2} (\partial_\mu \sigma)^2 -\dfrac{\alpha}{4} g^2
(\partial_\mu
\sigma \hat C_\nu - \partial_\nu \sigma \hat C_\mu)^2.
\label{slag2}
\eea
At this point one cannot miss the remarkable contrast between the
Lagrangians (\ref{smlag}) and (\ref{slag2}). They describe the
same theory, but obviously have totally different implications.
The Lagrangian (\ref{smlag}) describes a non-linear sigma model
which has no trace of gauge interaction. But the Lagrangian
(\ref{slag2}) clearly has the structure of gauge interaction which
allows us to relate the Skyrme theory to QCD.

Indeed we can derive the linearized Skyrme Lagrangian from QCD
\cite{cho01}. Consider $SU(2)$ QCD again. Introducing an
isotriplet unit vector field $\n$ which selects the ``Abelian''
direction (i.e., the color charge direction) at each space-time
point, we can decompose the gauge potential into the restricted
potential $\hat B_\mu$ and the valence potential $\vec X_\mu$
\cite{cho80,cho81},
\bea
& \vec{A}_\mu =A_\mu \n - \oneg \n\times\pro_\mu\n+\X_\mu\nonumber
= \hat B_\mu + \X_\mu, \nn\\
&  (\n^2 =1,~~~ \hat{n}\cdot\vec{X}_\mu=0),
\label{cdec}
\eea
where $A_\mu = \n\cdot \vec A_\mu$ is the ``electric'' potential.
Notice that the restricted potential is precisely the connection
which leaves $\n$ invariant under the parallel transport,
\bea
\D_\mu \n = \pro_\mu \n + g {\hat B}_\mu \times \n = 0.
\eea
Under the infinitesimal gauge transformation
\bea
\delta \n = - \vec \alpha \times \n  \,,\,\,\,\,
\delta \A_\mu = \oneg  D_\mu \vec \alpha,
\eea
one has
\bea
&&\delta A_\mu = \oneg \n \cdot \pro_\mu \valpha,\,\,\,\
\delta \hat B_\mu = \oneg \D_\mu \valpha  ,  \nn \\
&&\hspace{1.2cm}\delta \X_\mu = - \valpha \times \X_\mu  .
\eea
This shows that $\hat B_\mu$ by itself describes an $SU(2)$
connection which enjoys the full $SU(2)$ gauge degrees of freedom.
Furthermore $\vec X_\mu$ transforms covariantly under the gauge
transformation. Most importantly, the decomposition (\ref{cdec})
is gauge-independent. Once the color direction $\hn$ is selected
the decomposition follows automatically, independent of the choice
of a gauge.

Notice that the unit isotriplet $\hat{n}$ describes all
topological features of the original non-Abelian gauge potential.
Clearly the isolated singularities of $\hat{n}$ defines
$\pi_2(S^2)$ which describes the Wu-Yang monopole
\cite{cho80,cho81}. Besides, with the $S^3$ compactification of
$R^3$, $\hat{n}$ characterizes the Hopf invariant
$\pi_3(S^2)\simeq\pi_3(S^3)$ which classifies the topologically
distinct vacua and the instantons \cite{cho02,cho79}. The
importance of the decomposition (\ref{cdec}) has recently been
apprecisted by many authors in studying various aspects of QCD
\cite{fadd2,gies}. Furthermore in mathematics the decomposition
has been shown to play a crucial role in studying the geometry, in
particular the Deligne cohomology, of non-Abelian gauge theory
\cite{cho75,zucc}.

To understand the physical meaning of the decomposition
(\ref{cdec}) notice that the restricted potential $\hat{B}_\mu$
actually has a dual structure. Indeed the field strength made of
the restricted potential is decomposed as
\begin{eqnarray}
&\hat{B}_{\mu\nu}=\hat F_{\mu\nu}+ \hat H_{\mu\nu}
=(F_{\mu\nu}+ \tilde H_{\mu\nu}) \hn, \nonumber\\
&F_{\mu\nu}=\partial_\mu A_{\nu}-\partial_{\nu}A_\mu, \nn\\
&\tilde H_{\mu\nu}=-\dfrac{1}{g} \hn \cdot (\partial_\mu\hat{n}
\times\partial_\nu\hat{n})
=-\dfrac{1}{g} H_{\mu\nu} \nn\\
&=\partial_\mu \tilde C_\nu-\partial_\nu \tilde C_\mu,
\end{eqnarray}
where $\tilde C_\mu$ is the ``magnetic'' potential
\cite{cho80,cho81}. Notice that this magnetic potential in QCD is
identical (up to the normalization factor $-1/g$) to the magnetic
potential $C_\mu$ we have in Skyrme theory. This confirms that
$\hat C_\mu$ which appears in the Skyrme-Faddeev theory in
(\ref{ccon}) is precisely the magnetic potential which provides
the dual structure of QCD. This is an indication that the Skyrme
theory and QCD are dual to each other.

With (\ref{cdec}) we have
\bea
\vec F_{\mu\nu}=\hat B_{\mu \nu} + \D_\mu \X_\nu -
\D_\nu \X_\mu + g\X_\mu \times \X_\nu,
\eea
so that the Yang-Mills Lagrangian is expressed as
\bea
&{\cal L}=-\dfrac{1}{4} \vec F^2_{\mu \nu } \nn\\
&=-\dfrac{1}{4} {\hat B}_{\mu\nu}^2
-\dfrac{1}{4} ( \D_\mu \X_\nu -\D_\nu \X_\mu)^2 \nn\\
&-\dfrac{g}{2} {\hat B}_{\mu\nu}
\cdot (\X_\mu \times \X_\nu)
- \dfrac{g^2}{4} (\X_\mu \times \X_\nu)^2.
\eea
This tells that QCD can be viewed as a restricted gauge theory
made of the binding gluon $\hat B_\mu$ described by the restricted
potential, which has an additional valence gluon $\X_\mu$
described by the valence potential \cite{cho80,cho81}.

Now we can show that one can actually derive the linearized Skyrme
theory from a massive QCD. Suppose that the confinement mechanism
generates a mass term for the binding gluon. In this case the QCD
Lagrangian can be modified to a massive QCD
\bea
&{\cal L} \simeq -\dfrac{1}{4} {\hat B}_{\mu\nu}^2
-\dfrac{\mu^2}{2} \hat B_\mu^2
-\dfrac{1}{4} ( \D_\mu \X_\nu -\D_\nu \X_\mu)^2 \nn\\
&-\dfrac{g}{2} {\hat B}_{\mu\nu}
\cdot (\X_\mu \times \X_\nu)
- \dfrac{g^2}{4} (\X_\mu \times \X_\nu)^2.
\eea
Of course, this Lagrangian is too simple to describe the real
dynamical symmetry breaking in QCD. But notice that, once the
confinement sets in, the insertion of the mass term can be
justified. Now the above Lagrangian, in the absence of $A_\mu$ and
$\vec X_\mu$, reduces exactly to the Skyrme-Faddeev Lagrangian
(\ref{sflag}). Furthermore, with
\bea
\X_\mu = f_1 \partial_\mu \n + f_2 \n \times \partial_\mu \n,
\eea
the Lagrangian is expressed as
\bea
&{\cal L} \simeq -\dfrac{1}{4} \big[ F_{\mu \nu} +(1-g\phi^* \phi)
\tilde H_{\mu \nu} \big] ^2 \nn\\
&+\dfrac{ig}{2} (D_\mu \phi)^* (D_\nu \phi) \tilde H_{\mu \nu}
-\dfrac{g^2}{4} \big| D_\mu \phi \hat C_\nu
-D_\nu \phi \hat C_\mu \big|^2 \nn\\
&-\dfrac{\mu^2}{2} \big(A_\mu^2 + \hat C_\mu^2 \big),
\eea
where
\bea
\phi = f_1 + if_2,
~~~~~D_\mu \phi = (\partial_\mu + i g A_\mu) \phi. \nn
\eea
So, with
\bea
\partial_\mu \phi = 0,~~~~~A_\mu = \partial_\mu \sigma,
\eea
we have
\bea
&{\cal L} \simeq -\dfrac{(1-g \phi^* \phi)^2}{4} g^2 \hat
H_{\mu\nu}^2
-\dfrac{\mu^2}{2} g^2 \hat C_\mu^2 \nn\\
&-\dfrac{\mu^2}{2} (\partial_\mu \sigma)^2 -\dfrac{\phi^* \phi}{4}
g^2 (\partial_\mu \sigma \hat C_\mu -\partial_\nu \sigma \hat
C_\nu)^2.
\label{mqcd}
\eea
Now, with
\bea
\alpha = \phi^* \phi = (1-g \phi^* \phi)^2,
\eea
the Lagrangian (\ref{mqcd}) becomes nothing but the linearized
Skyrme Lagrangian (\ref{slag2}). So, if one wishes, one can
actually derive the linearized Skyrme theory from QCD
\cite{cho01,cho3}. This confirms the deep connection between the
Skyrme theory and QCD. More importantly this demonstrates that the
Skyrme theory describes the chromomagnetic dynamics, not the
chromoelectric dynamics, of QCD. This tells that the two theories
are dual to each other.

The new interpretation is completely orthogonal to the popular
view. Nevertheless two views are not necessarily contradictory to
each other, in the sense that both agrees that the Skyrme theory
is an effective theory of strong interaction in which the
confinement is manifest already at the classical level. Indeed we
believe that the new interpretation should be viewed complementary
to the popular view, which can allow us a deeper understanding of
the Skyrme theory.

\section{Chromoelectric Knot in QCD}

Based on the fact that the Skyrme theory is closely related
to QCD, Faddeev and Niemi predicted a topological knot
in QCD \cite{fadd2}. According to the dual
picture, the Faddeev-Niemi knot should be interpreted as a
chromomagnetic knot made of twisted magnetic flux, which can
not exist in QCD because QCD confines chromoelectric
(not chromomagnetic) flux. But the dual picture implies
the existence of a chromoelectric knot in QCD which is dual to
the chromomagnetic Faddeev-Niemi knot \cite{plb05}. In other
words it could have a knot made of twisted color electric flux
ring. This, of course, is a bold conjecture. Proving this
conjecture within the framework of QCD will require a detailed
knowledge of the confinement mechanism. Nevertheless from the
physical point of view there is no reason why such an object can
not exist, because one can easily construct such object simply by
twisting a $g\bar g$ flux and smoothly bending and connecting both
ends. Assuming the existence one may estimate the mass of the
lightest electric knot. In this case one may identify
$\Lambda_{QCD}$ as $({\sqrt \alpha}\mu)_{QCD}$, because this is
the only scale one has in QCD. Now, with \cite{pesk}
\bea
\Lambda_{QCD} \simeq ({\sqrt \alpha}\mu)_{QCD} \simeq 200~MeV,
\label{sqcd}
\eea
one can easily estimate the mass of the lightest electric knot.
From (\ref{skke}) we expect \cite{plb05}
\bea
M_K \simeq 305.8~\Lambda_{QCD} \simeq 61.2~GeV.
\label{qcdke}
\eea
It would certainly be very interesting to search for such exotic
glueball experimentally.

At this point one might wonder how the small $\Lambda_{QCD}$ can
produce such a large mass. To understand this notice that in
Skyrme theory we have a large knot energy in spite of the fact
that the theory has only one small mass scale ${\sqrt \alpha}\mu$.
The reason is because of the large volume of the knot. Here we
have exactly the same situation. To see this notice that the
radius of the vortex ring of the lightest Faddeev-Niemi knot is
about 3.6 times larger than the radius of vortex. So assuming that
the radius of vortex of the chromoelectric knot in QCD is about
$1/\Lambda_{QCD}$, we expect the volume the knot to be about
$7.2~\pi^2 /\Lambda_{QCD}^{3}$. Now assuming that the energy
density $\rho$ of the knot is comparable to that of typical
low-lying hadrons, we expect $\rho$ roughly to be of the order of
$\Lambda_{QCD}^{3} M_p$ (where $M_p$ is the proton mass). From
this we expect the energy of the knot to be about $66.7~GeV$,
which tells that our estimate (\ref{qcdke}) is quite reasonable.
This explains how the small $\Lambda_{QCD}$ can produce such a
large knot mass. As importantly this implies that, as far as the
energy density is concerned, our knot is not much different from
ordinary hadrons, in spite of its large mass.

Classically the chromoelectric knot must be stable, because it has
both topological and dynamical stability. But this does not
guarantee the physical stability of the knot, because in QCD we
have other fields, the quarks and gluons, which could destabilize
the knot. For example, the knot can be cut and decay to $g \bar g$
pairs and thus to lowlying hadrons. Indeed, just as the electric
background in QED produces electron-positron pairs \cite{sch}, the
chormoelectric background in QCD becomes unstable and decays to
gluon pairs \cite{cho04}. This tells that the knot must have the
quantum instability.

We could estimate the decay width of the knot from the one-loop
effective action of QCD, because the effective action tells us
what is the pair production probability of the gluons in
chromoelectric background. In the presence of pure electric or
pure magnetic background the one-loop effective action of $SU(2)$
QCD is given by \cite{cho02,cho04}
\bea
{\cal L}_{eff}=\left\{\begin{array}{ll}-\dfrac{a^2}{2g^2}
-\dfrac{11a^2}{48\pi^2}
(\ln \dfrac{a}{\mu^2}-c),~~~~~b=0 \\
~\dfrac{b^2}{2g^2} +\dfrac{11b^2}{48\pi^2}
(\ln \dfrac{b}{\mu^2}-c) \\
-i\dfrac{11b^2}{96\pi}, ~~~~~~~~~~~~~~~~~~~~~~~~~~~~a=0
\end{array}\right.
\label{ceaab}
\eea
where $c$ is a (subtraction-dependent) constant and
\bea
&a = \dfrac{g}{2} \sqrt {\sqrt {G^4 + (G \tilde G)^2} + G^2}, \nn\\
&b = \dfrac{g}{2} \sqrt {\sqrt {G^4 + (G \tilde G)^2} - G^2}, \nn\\
&G_{\mu\nu} = F_{\mu\nu} + H_{\mu\nu}. \nn
\eea
Notice that $a=gH$ and $b=0$ represent a pure magnetic background,
and $a=0$ and $b=gE$ represent a pure electric background.
According to the effective action the chromoelectric background is
unstable and decays to $g \bar g$, with the probability $11g^2
E^2/96\pi$ per unit volume per unit time \cite{cho02,cho04}. So
assuming that the knot is made of $g \bar g$ flux ring of
thickness $1/\Lambda_{QCD}$ we can estimate the average electric
field of the knot to be
\bea
\bar E \simeq \dfrac{g\Lambda_{QCD}^2}{\pi}.
\eea
Also the lightest knot whose radius is about $3.6$ times the
thickness of the knot has the volume $7.2~\pi^2
/\Lambda_{QCD}^{3}$.
From this we can estimate the decay width $\Gamma$
of the knot \cite{plb05}
\bea
&\Gamma \simeq \dfrac{11g^2}{96\pi}
\Big(\dfrac{g\Lambda_{QCD}^2}{\pi} \Big)^2
\times \dfrac{7.2~\pi^2}{\Lambda_{QCD}^3} \nn\\
&\simeq 41~\alpha_s^2~\Lambda_{QCD}.
\label{dwk}
\eea
Obviously the decay width depends on the volume of the knot which
is large. But more importantly it depends on $\alpha_s$ which is
running. This is because the decay process is a quantum process.
So we have to decide what value of $\alpha_s$ we have to use.
There are two extremes, $\alpha_s(M_p) \simeq 1$ and
$\alpha_s(M_k) \simeq 0.13$ \cite{pesk}. We prefer $\alpha_s
\simeq 1$. This is because, as far as the energy density is
concerned, the knot is like ordinary hadrons in spite of the large
mass. With $\alpha_s \simeq 1$ we have $\Gamma \simeq 8.2~GeV$. So
we expect a knot glueball of mass about $60~GeV$ which has a decay
width about $8~GeV$. Of course this is a rough estimate, but this
implies that the chromoelectric knot can have a typical hadronic
decay. In this connection we remark that, had we used
$\alpha_s(M_k) \simeq 0.13$, the decay width would have been
around $0.14~GeV$ which is absurd. In the presence of quarks, a
similar knot made of a twisted $q\bar q$ flux could also exist in
QCD.

The above exercise teaches us an important lesson. It tells that
the classical stability of a topological soliton does not
guarantee the physical stability. It can decay through a quantum
process. The decay of the chromoelectric knot demonstrates this
fact.

At this point one may ask what is the difference between the
Faddeev-Niemi knot in Skyrme theory and our knot in QCD. Clearly
both theories advocates the existence of a hadronic knot soliton
in strong interaction. If one believes in the popular view
of Skyrme theory as an effective theory of
strong interaction, one must expect a hadronic knot around
$5~GeV$. But if one believes in the dual picture, one should
expect a hadronic knot around $60$ GeV. Experiment can tell which
view is more realistic. A nice feature of the dual picture is that
one can identify not just the decay mechanism, but actually
calculate the decay width of the knot within the framework of QCD.
This is impossible for the Faddeev-Niemi knot in Skyrme theory.

\section{Skyrme Theory and Condensed Matter Physics}

The Skyrme theory allows another unexpected interpretation. With
the built-in Meissner effect, the (helical) baby skyrmion looks
very much like the magnetic vortex in superconductors. This
strongly suggests that the Skyrme theory could be related to a
condensed matter physics at low energy \cite{cho1,cho2}. Now we
argue that the Faddeev-Niemi Lagrangian (\ref{sflag}) can actually
be viewed to describe a $CP^1$ model which describes a
two-component superfluid.

To see this let $\phi$ be a complex doublet which describes a
two-component Bose-Einstein condensate (BEC)
\bea
\phi = \dfrac {1}{\sqrt 2} \rho \xi , ~~~~~(\xi^\dag \xi = 1)
\eea
and consider the following ``gauged" Gross-Pitaevskii type
Lagrangian \cite{cho1}
\bea
&{\cal L} = - |D_\mu \phi|^2 - \dfrac {\lambda}{2}
\big(\phi^\dag \phi -\dfrac{\mu^2}{\lambda} \big)^2
- \dfrac {1}{4} F_{\mu \nu} ^2,
\label{beclag}
\eea
where $D_\mu =  \pro_\mu + i g A_\mu$, $\mu^2$ and $\lambda$ are
the coupling constants. Of course, since we are interested in a
neutral condensate, we identify the potential $A_\mu$ with the
velocity field of $\xi$ \cite{cho1}
\bea
g A_\mu = -i \xi^\dag \pro_\mu \xi.
\label{vpot}
\eea
With this the Lagrangian (\ref{beclag}) is reduced to
\bea
&{\cal L} = -\dfrac {1}{2} (\pro_\mu \rho)^2 - \dfrac {\rho^2}{2}
\Big(|\pro_\mu
\xi |^2 - |\xi^\dag \pro_\mu \xi|^2 \Big) \nn\\
&-\dfrac{\lambda}{8} \Big(\rho^2 - \rho_0^2 \Big)^2 + \dfrac {1}{4
g^2} (\pro_\mu \xi^\dag \pro_\nu \xi - \pro_\nu \xi^\dag \pro_\mu
\xi)^2,
\label{beclag1}
\eea
where
\bea
\rho_0^2=\dfrac{2\mu^2}{\lambda}. \nn
\label{vac}
\eea
Notice that here the gauge field strength $F_{\mu\nu}$ is replaced
by the non-vanishing vorticity of the velocity field (\ref{vpot}),
but the Lagrangian still retains the $U(1)$ gauge symmetry of
(\ref{beclag}).

From the Lagrangian we have the following equation of motion
\bea
& \pro^2 \rho - \Big(|\pro_\mu \xi |^2 - |\xi^\dag \pro_\mu \xi|^2
\Big)
\rho = \dfrac{\lambda}{2} (\rho^2 - \rho_0^2) \rho,\nn \\
&\Big\{(\pro^2 - \xi^\dag \pro^2 \xi) + 2 \Big(\dfrac {\pro_\mu
\rho}{\rho} - \xi^\dag \pro_\mu\xi \nn\\
&+ \dfrac {1}{g^2 \rho^2} \pro_\alpha (\pro_\mu
\xi^\dag \pro_\alpha \xi - \pro_\alpha \xi^\dag \pro_\mu \xi) \Big)
(\pro_\mu - \xi^\dag \pro_\mu \xi) \Big\} \xi \nn\\
&= 0.
\label{beceq1}
\eea
But remarkably, with
\bea
\n = \xi^\dag \vec \sigma \xi,
\label{ndef}
\eea
we have
\bea
& (\pro_\mu \hn)^2 = 4 (|\pro_\mu \xi|^2
- |\xi^\dag \pro_\mu \xi|^2) \nn\\
&H_{\mu\nu} = \hn \cdot (\pro_\mu \hn \times \pro_\nu \hn) = -2i
(\pro_\mu \xi^\dag \pro_\nu \xi
- \pro_\nu \xi^\dag \pro_\mu \xi ) \nn\\
&= \pro_\mu C_\nu - \pro_\nu C_\mu.
\label{fmn}
\eea
This tells that the velocity potential (\ref{vpot}) plays the role
of the magnetic potential $C_\mu$ in (\ref{sfeq}) of Skyrme
theory.

With (\ref{fmn}) the Lagrangian (\ref{beclag1}) is reduced to the
following $CP^1$ Lagrangian
\bea
&{\cal L} = -\dfrac{1}{2}(\pro_\mu \rho)^2 -\dfrac {\rho^2}{2}
(\pro_\mu \hn)^2
-\dfrac{\lambda}{8} (\rho^2-\rho_0^2)^2 \nn\\
&-\dfrac {1}{16g^2} (\pro_\mu \hn \times \pro_\nu \hn)^2.
\label{beclag3}
\eea
Furthermore the equation (\ref{beceq1}) can be put into the form
\bea &\pro^2 \rho - \dfrac{1}{4} (\pro_i
\n)^2 \rho = \dfrac{\lambda}{2} (\rho^2 - \rho_0^2)
\rho, \nn \\
&\n \times \pro^2 \n + 2 \dfrac{\pro_i \rho}{\rho} \n \times
\pro_i
\n + \dfrac{1}{g^2 \rho^2} \pro_i H_{ij} \pro_j \n = 0.
\label{beceq2}
\eea
The reason why we can express (\ref{beceq1}) completely in terms
of $\hn$ (and $\rho$) is that the Abelian gauge invariance of
(\ref{beclag}) effectively reduces the target space of $\xi$ to
the gauge orbit space $S^2 = S^3/S^1$, which is identical to the
target space of $\n$.

This analysis clearly shows that the above theory of two-component
BEC is closely related to the Skyrme theory. In fact, in the
vacuum
\bea
\rho^2=\rho_0^2,
\label{becvac}
\eea
the Lagrangian ie reduced to the Skyrme-Faddeev Lagrangian
\bea
&{\cal L} = -\dfrac {\rho_0^2}{2} \Big(|\pro_\mu
\xi |^2 - |\xi^\dag \pro_\mu \xi|^2 \Big) \nn\\
&- \dfrac {1}{4g^2} (\pro_\mu \xi^\dag \pro_\nu \xi
- \pro_\nu \xi^\dag \pro_\mu \xi)^2 \nn\\
&=-\dfrac {\rho_0^2}{2} (\pro_\mu \hn)^2 -\dfrac {1}{16g^2}
(\pro_\mu \hn \times \pro_\nu \hn)^2.
\label{beclag2}
\eea
This tells that three Lagrangians (\ref{sflag}), (\ref{qcdlag}),
and (\ref{beclag2}) are all identical to each other, which
confirms that the Skyrme theory and the theory of two-component
superfluid indeed have an important overlap. This implies that the
Skyrme-Faddeev theory could also be regarded as a theory of
two-component superfluid.

The above analysis also reveals two important facts. First, it
shows that the Skyrme theory has a $U(1)$ gauge symmetry. This is
evident from the fact that the two-form $H_{\mu\nu}$ admits the
gauge potential given by the velocity field (\ref{vpot}) of $\xi$.
Actually the gauge symmetry really originates from the little
group of $\hn$, the arbitrary $U(1)$ rotation which leaves $\hn$
(and thus the Skyrme Lagrangian) invariant. As we have seen, it is
this gauge symmetry which have allowed us to establish the
existence of the Meissner effect in Skyrme theory. Secondly, it
provides a new meaning to $H_{\mu\nu}$. The two-form now describes
the vorticity of the velocity field of the superfluid $\xi$. In
other words, the non-linear Skyrme interaction can be interpreted
as the vorticity interaction in superfluids. It has been
well-known that the vorticity plays an important role in
superfluids \cite{ho}. But creating a vorticity in superfluid
costs energy. So it makes a perfect sense to include the vorticity
interaction in the theory of superfluids. And the Skyrme-Faddeev
Lagrangian naturally contains this interaction.

Furthermore the above analysis makes it clear that the above gauge
theory of two-component BEC has a knot solution very similar to
the Faddeev-Niemi knot \cite{cho1,cho2}. Indeed the following knot
ansatz for two-component BEC
\bea
&\phi=\dfrac{1}{\sqrt 2} \rho (\eta, \gamma) \xi, \nn\\
&\xi=\Bigg(\matrix{\cos \dfrac{f(\eta,\gamma )}{2} \exp
(-im\varphi ) \cr
\sin \dfrac{f(\eta ,\gamma)}{2} \exp (in\beta (\eta,\gamma))} \Bigg),
\label{bkans}
\eea
gives us
\bea
&\hn= \xi^{\dagger} \vec \sigma \xi =\Bigg(\matrix{\sin f
\cos (n\beta+m\varphi) \cr
\sin f \sin (n\beta + m\varphi) \cr
\cos f } \Bigg),  \nn\\
&C_\mu = -2i \xi^{\dagger} \partial _\mu \xi \nn\\
&= n(\cos f -1) \partial _\mu \beta +m(\cos f +1)\partial _\mu
\varphi. \nn
\eea
This is identical to the knot ansatz (\ref{skkans}) for the
Faddeev-Niemi knot. Indeed with the ansatz (\ref{bkans}) we can
obtain a knot solution very similar to the Faddeev-Niemi knot
\cite{cho1,cho2}. The only difference between the two knots is
that the one in BEC has a dressing of an extra scalar field $\rho$
which represents the degree of the condensation. This implies that
under a proper circumstance, the condensed matter physics can
allow a knot similar to the Faddeev-Niemi knot.

Clearly the knot in two component BEC describes a vorticity knot.
But here the complex doublet $\xi$ provides the knot topology,
because it defines the mapping $\pi_3(S^3)$ from the compactified
space $S^3$ to the target space $S^3$ of the unit doublet. But
with the Hopf fibering of $S^3$ to $S^2 \times S^1$, we have
$\pi_3(S^3) \simeq \pi_3(S^2)$. So two mappings $\pi_3(S^2)$
defined by $\hn$ and $\pi_3(S^3)$ defined by $\xi$ describe an
identical knot topology. Indeed in terms of the complex doublet
$\xi$ the knot quantum number is given by
\bea
&Q_k = \dfrac {1}{4\pi^2} \int \epsilon_{ijk} \xi^{\dagger}
\partial_i \xi ( \partial_j \xi^{\dagger}
\partial_k \xi ) d^3 x  \nn\\
&= \dfrac{1}{32\pi^2} \int \epsilon_{ijk} C_i H_{jk} d^3x,
\label{bkqn}
\eea
which is identical to the knot quantum number of Skyrme theory.

The fact that the knot quantum number of Two-component BEC can be
described by $\pi_3(S^3)$ has led to a confusing statement in the
literature that the knot can be identified as a skyrmion, because
the baryon quantum number in Skyrme theory is also described by
$\pi_3(S^3)$ \cite{batt}. But we emphasize that this is a
misleading statement, because $\pi_3(S^3)$ of (\ref{bkqn}) in BEC
is different from the one which defines the baryon number
(\ref{bn}) in Skyrme theory. In both cases the target space $S^3
\simeq S^2 \times S^1$ has the Hopf fiber $S^1$. But for the
$\pi_3(S^3)$ of the knot in BEC the hidden $U(1)$ gauge group
constitutes the fiber $S^1$, so that $\pi_3(S^3)$ defined by $\xi$
actually reduces to $\pi_3(S^2)$. On the other hand, for the
$\pi_3(S^3)$ of the skyrmion in (\ref{bn}) the massless scalar
field $\omega$ describes the fiber $S^1$. And the non-trivial
$\omega$ forbids us to reduce this $\pi_3(S^3)$ to $\pi_3(S^2)$.
So the two $\pi_3(S^3)$ actually describe different topology. This
tells that it is misleading to call the knot in BEC a skyrmion.

The above analysis also makes it clear that alternatively the
Faddeev-Niemi knot can also be viewed as a two quantized vorticity
rings linked together in a two-component superfluid, whose linking
number becomes the knot quantum number \cite{cho1}.

\section{Discussion}

The Skyrme theory has rich topological structures. It has the
Wu-Yang type monopole which has infinite energy, the skyrmion
which can be viewed as a finite energy dressed monopole, the baby
skyrmion which can be viewed as an infinitely long magnetic flux
of a monopole-antimonopole pair, and the Faddeev-niemi knot which
can be viewed as a twisted magnetic vortex ring made of a helical
baby skyrmion. Both the monopole and the baby skyrmion have the
topology $\pi_2(S^2)$. But the same topology describes physically
different mapping, and thus physically different objects.
Similarly the skyrmion and the knot have different topology. But
the topologies of these objects are closely related. With the Hopf
fibering $S^3=S^2\times S^1$ the monopole topology $\pi_2(S^2)$
can naturally be extended to $\pi_3(S^3)$ of the skyrmion
topology. Similarly $\pi_2(S^2)$ of the baby skyrmion is extended
to $\pi_3(S^2)$ of the helical baby skyrmion (and the knot)
topology.

In this paper we have presented the numerical solution of the knot
in Skyrme theory. This was made possible with a consistent knot
ansatz. The numerical result confirms that the knot is indeed a
twisted magnetic vortex ring made of an helical baby skyrmion. In
fact we have shown that the knot can be viewed as two magnetic
flux rings linked together, whose linking number is fixed by the
knot quantum number. This confirms that the knot has a dynamical
manifestation of knot topology which assures the dynamical
stability.

But what is really remarkable is that the Skyrme theory itself has
many different faces. The Skyrme theory has always been associated
to nuclear and/or high energy physics at $GeV$ scale. This
traditional view is based on the fact that the theory can be put
into the form of a non-linear sigma model which describes the
flavor dynamics. As we have shown, however, our analysis tells
that theory can also be interpreted as an effective theory of
chromomagnetic dynamics which is dual to QCD. More precisely it
can be viewed as an effective theory of confinement with a
built-in Meissner effect. This view is completely orthogonal to,
but not inconsistent with, the traditional view. Both shows that
the Skyrme theory can be interpreted as an effective theory of
strong interaction. In this sense the new interpretation is
complementary to the traditional interpretation.

Both views predict the existence of a non-baryonic
topological knot in QCD, but
they predict different knot. In the traditional view the mass
of the lightest knot is supposed to be around $5~GeV$,
but in the dual picture the mass of such knot should be around
$60~GeV$. Experimentally one could tell which is a better view
simply by measuring (assuming the existence) the mass of such
exotic knot.

The Skyrme theory has another surprising face. The Meissner effect
in Skyrme theory suggests that the theory could be interpreted as
a theory of condensed matter. In this paper we have argued that
the Skyrme-Faddeev Lagrangian (\ref{sflag}) could be understood to
describe a theory of two-component superfluid or two-component
Bose-Einstein condensate. In particular, the Faddeev-Niemi knot
could be viewed as a vorticity vortex ring in these condensed
matters \cite{cho1,cho2}. This implies that the theory could
actually describe an interesting low energy physics in a
completely different environment at $eV$ scale, in two-component
condensed matters.

If this view is correct, one could actually construct the
Faddeev-Niemi knot (or a similar one) in laboratories, in
particular in two-component superfluids and/or two-gap
superconductors \cite{cho1,cho2}. If so, the challenge now is to
confirm the existence of the topological knot experimentally in
these condensed matters. Constructing such knot may be a tricky
task at present moment, but we hope that such a knot could be
constructed in laboratories in the near future.

\acknowledgments

~~~One of us (YMC) thanks G. Sterman for the kind hospitality
during his visit to Institute for Theoretical Physics. The work is
supported in part by the ABRL Program of Korea Science and
Engineering Foundation (R14-2003-012-01002-0) and by the BK21
Project of the Ministry of Education.


\end{document}